\documentclass[11pt,a4paper]{article}
\pdfoutput=1

\usepackage{jheppub}

\usepackage{amsmath}
\usepackage{bbm}
\usepackage{float}
\usepackage{graphicx}	
\usepackage{hyperref}
\usepackage{mathtools}
\usepackage{cancel}

\usepackage[normalem]{ulem}

\usepackage{multirow}
\usepackage{rotating}
\usepackage{subcaption}

\def\lsim{\raise0.3ex\hbox{$\;<$\kern-0.75em\raise-1.1ex
\hbox{$\sim\;$}}}
\def\gsim{\raise0.3ex\hbox{$\;>$\kern-0.75em\raise-1.1ex
\hbox{$\sim\;$}}}

\def\thetitle{ 
Perturbing neutrino oscillations around the solar resonance \\
}
\title{\thetitle}
\hypersetup{pdftitle={\thetitle}}
\author{Ivan Martinez-Soler$^{a,b,c}$}
\author{Hisakazu Minakata$^{d,e}$}

\affiliation{
  $^a$Theoretical Physics Department, Fermi National Accelerator Laboratory, P.O. Box 500, Batavia IL 60510, USA \\
  $^b$Department of Physics and Astronomy, Northwestern University, Evanston, IL 60208, USA \\
  $^c$Colegio de F\'isica Fundamental e Interdisciplinaria de las Am\'ericas (COFI), 254 Norzagaray street, San Juan, Puerto Rico 00901 \\
  $^d$Center for Neutrino Physics, Department of Physics, Virginia Tech, Blacksburg, Virginia 24061, USA \\
  $^e$Research Center for Cosmic Neutrinos, Institute for Cosmic Ray Research, University of Tokyo, Kashiwa, Chiba 277-8582, Japan \\ 
}

\emailAdd{ivan.martinezsoler@northwestern.edu}
\emailAdd{minakata71@vt.edu}
\date{\today}

\abstract{ 
Atmospheric neutrinos at low energies, $E \lsim 500$ MeV, is known to be a rich source of information of lepton mixing parameters. We formulate a simple perturbative framework to elucidate the characteristic features of neutrino oscillation at around the solar-scale enhancement due to the matter effect. The clearest message we could extract from our perturbation theory is that CP violation in the appearance oscillation probability is large, a factor of $\sim 10$ times larger than CP violation at around the atmospheric-scale oscillation maximum. Underlying mechanism for it is that one of the suppression factors on the CP phase dependent terms due to smallness of $\Delta m^2_{21} / \Delta m^2_{31}$ are dynamically lifted by the solar-scale enhancement. Our framework has a unique feature as a perturbation theory in which large $\Delta m^2_{31}$ term outside the key 1-2 sector for the solar-scale resonance does not yield sizeable corrections. On the contrary, the larger the $\Delta m^2_{31}$, the smaller the higher order corrections. 
}

\begin{document} % JHEP 

\begin{flushright}
FERMILAB-PUB-19-149-T,
NUHEP-TH/19-05
\end{flushright}

\maketitle

\section{Introduction}
\label{sec:introduction}

Physics of neutrinos has been blossomed in the last 20 years after the discovery of neutrino oscillation \cite{Fukuda:1994mc,Fukuda:1998mi,Eguchi:2002dm,Cleveland:1998nv,Ahmad:2002jz}, which implies nonzero masses of neutrinos and existence of lepton flavor mixing \cite{Maki:1962mu}. 
It is interesting to observe that pinning down these evidences of neutrino mass required understanding of the matter effect to a different degree of importance. While no or limited importance of the matter effect in KamLAND reactor $\bar{\nu}_{e}$ measurement and the atmospheric neutrino observation in Super-Kamiokande, the theory of matter enhanced flavor conversion \cite{Mikheev:1986gs,Wolfenstein:1977ue} provided the key to understand the results of solar neutrino experiments pioneered by Davis {\it et al.}, in particular, the charged current/neutral current (CC/NC) ratio measured by SNO.

Another important feature of neutrino oscillation we see is the existence of two distance scales, $\ell_{ \text{atm} } \equiv \frac{ 4\pi E}{ \Delta m^2_{31} } \sim 100$ km and $\ell_{ \text{solar} } \equiv \frac{ 4\pi E}{ \Delta m^2_{21} } \sim 3000$ km at $E=100$ MeV,\footnote{
%%%%%%%%%%%%% footnote %%%%%%%%%%%%%%
$\ell_{ \text{atm} } = 103$ km and $\ell_{ \text{solar} } = 3.30 \times 10^3$ km 
if we use $\Delta m^2_{31}=2.4 \times 10^{-3}~\mbox{eV}^2$, $\Delta m^2_{21}=7.5 \times 10^{-5}~\mbox{eV}^2$, $E=100~\mbox{MeV}$. 
}
which so much differ from each other because of the hierarchy of mass squared differences $\Delta m^2_{31} / \Delta m^2_{21} \simeq 30$ in our three-generation world of fundamental leptons. It is interesting to observe that both of them easily fit into our mother planet. This is shown in the left panel of figure~\ref{fig:Pmue-color-grade} in section~\ref{sec:how-accurate}, in which the equi-oscillation probability contours of $\nu_{\mu} \rightarrow \nu_{e}$ channel is plotted in space of neutrino energy $E$ and baseline $L$. It is notable that both of the two matter-effect enhanced regions, or ``resonances'', are clearly visible and equally prominent. 
Again, it is intriguing to see that the matter density of the earth is designed in such a way that they both live inside the plot as in the left panel of figure~\ref{fig:Pmue-color-grade}. 

Since the discovery era, all the mixing angles and the two $\Delta m^2$ have been determined by extensive efforts by the experimentalists, see for example, refs.~\cite{Gando:2013nba,Abe:2018wpn,Abe:2017bay,NOvA:2018gge,Jiang:2019xwn,Adey:2018zwh,Bak:2018ydk,Nakano} for a very limited list. The latest solar neutrino combined results are given in Fig.~13.2 of ref.~\cite{Nakano}. 
Then, what is left inside $\nu$SM by now is to measure CP violating phase $\delta$ and determination of the neutrino mass ordering. It appears that the strategies for determination of ``what is left'' are well formulated. The tantalizing hint for $\delta \sim \frac{3\pi}{2}$ from T2K experiment \cite{Okumura-ICRR-seminar} may be confirmed at higher confidence level (CL). The next-generation long-baseline (LBL) accelerator neutrino experiments, T2HK and DUNE \cite{Abe:2015zbg,Acciarri:2015uup}, are designed to perform definitive measurement of $\delta$. Determination of the neutrino mass pattern can be carried out by utilizing the earth matter effect\footnote{
%%%%%%%%%%%%% footnote %%%%%%%%%%%%%%
A simple understanding of why it can be done in this way is to note the approximate degeneracy in the oscillation probability in vacuum under the transformations $\Delta m^2_{31} \rightarrow - \Delta m^2_{31}$ and $\delta \rightarrow \pi - \delta$ is broken by the matter potential \cite{Minakata:2001qm}. Similarly, the interplay between the effects of $\delta$, the matter potential, and the different mass orderings is summarized in a compact manner by the bi-probability plot introduced in \cite{Minakata:2001qm}. 
}
by the same experiments, or for T2HK possibly by an extended version T2HKK \cite{Abe:2016ero}. It can also be done by atmospheric neutrino observation by them as well as by IceCube-Gen2/PINGU and/or KM3NeT-ORCA \cite{TheIceCube-Gen2:2016cap,Adrian-Martinez:2016zzs}.\footnote{
%%%%%%%%%%%%%% footnote %%%%%%%%%%%%%%
We are aware that the point of view described here may be too much simplified. See brief comments in section~\ref{sec:conclusion} for partially appending to the biased view on how to determine the remaining unknowns.
} 

It may be surprising to see that not only the ongoing and the future LBL experiments but also all the remaining ones, except for JUNO \cite{An:2015jdp}, 
rely on measurement at around the atmospheric oscillation maximum, $\frac{ \Delta m^2_{31} L }{ 4E } \sim \frac{\pi}{2}$. Given the feature that the solar oscillation maximum and the solar resonance are as prominent as the atmospheric ones, it must be more natural to pursue the line of solar-scale oscillation physics for the purpose of mixing parameter measurement. Whether right or wrong, it defines our physics motivation leading to the work described in this paper. 

It is the purpose of this paper to explore physics and possible advantage of measurement at the solar resonance affected region or around the solar oscillation maximum, for short, the region of ``solar-scale oscillation''.\footnote{
%%%%%%%%%%%%%% footnote %%%%%%%%%%%%%%
The essence of this paper was presented in Workshop for Atmospheric Neutrino Production, March 20-22, 2019, Nagoya University, Nagoya  \cite{Minakata-talk}.
}
We do it under a hope that the discovery potential in such region will be eventually fully explored experimentally. We must note, however, that this is our first attempt with ``pedestrian spirit'', in which we try to understand the characteristic features of neutrino oscillation at a qualitative level. Therefore, the readers must be aware that even when we address utility of solar-scale oscillation physics in the context of mixing parameter measurement, our statements will only be at a qualitative level. 

We happily note, with full respect to the foregoing works, that such an approach to solar-scale oscillation physics are neither new nor unique. Notably, the extensive foregoing studies by Smirnov and the collaborators on this subject must be mentioned \cite{Peres:2003wd,Peres:2009xe,Akhmedov:2008qt,Razzaque:2014vba}. Yet, we may still miss many other relevant works. The authors of refs.~\cite{Peres:2003wd,Peres:2009xe,Akhmedov:2008qt,Razzaque:2014vba} presented detailed studies of atmospheric neutrinos at low energies, and estimated, for example, how large is the effect of CP phase. Despite these existing massive works, we believe that physics of neutrino oscillations at around the solar oscillation maximum, in a broader sense, has a sufficient importance to be explored further. 
In fact, there are new ongoing researches toward this direction. Detection of low-energy atmospheric neutrinos by the large liquid scintillator detector is under investigation in the JUNO group with a merit of higher light output at low energies \cite{Mari-etal-Nutele2019}. The similar low energy region is studied with a large liquid Ar TPC, and the sensitivity to CP phase is evaluated in the context of DUNE \cite{Kelly:2019itm}.

In this paper, toward understanding physics of the solar-scale oscillation, we formulate a perturbative framework by expanding around the solar-scale enhancement. More specifically, we aim at constructing perturbation theory of neutrino oscillation by targeting the region $E \simeq \mbox{a few} \times 100$ MeV and $L \simeq \mbox{a few} \times 1000$ km. In a sense it is a continuation of the proposal of low energy $\sim$100 MeV conventional superbeam to measure $\delta$~\cite{Minakata:2000ee}, the cleanest place for CP measurement, but in a new unexplored territory in that work. 

What is the most interesting outcome of our perturbative treatment? We will show that CP violation is large at around the solar resonance, a factor of $\sim 10$ larger than those expected at the conventional LBL neutrino experiments such as T2HK and DUNE. As is well known, the effect of CP phase $\delta$, not only CP-odd but also CP-even effects, receives the two suppression factors, $\epsilon \equiv \Delta m^2_{21} / \Delta m^2_{31}$ and the product of angles $c_{23} s_{23} s_{13} c_{12} s_{12}$ \cite{Asano:2011nj}. The secret of the factor $\sim$10 enhancement, which gives us a definite advantage in looking for the CP phase effect, is that the former $\epsilon$ suppression goes away due to the solar-scale enhancement.
This property makes atmospheric neutrinos at low energies, say $E=100 - 500$ MeV, particularly interesting target for hunting lepton CP violation. We believe that our framework also has an interesting novel feature as a perturbation theory of neutrino oscillation.\footnote{
%%%%%%%%%%%%%% footnote %%%%%%%%%%%%%%%
Utility of perturbation theory of neutrino oscillation is, however, often an issue of controversy. Since this is a quite involved subject to discuss here in Introduction, we defer it to the next section~\ref{sec:utility-of-Ptheory} which is devoted to this specific topics. 
}
It will be explained in section~\ref{sec:what-is-new}.
We construct our perturbative framework, the ``solar resonance perturbation theory'', in section~\ref{sec:formulation}. The formulas for the neutrino oscillation probability is derived and their properties are elucidated in section~\ref{sec:oscillation-probability-symmetry}. Numerical accuracy of the probability formula is examined in section~\ref{sec:how-accurate}, which leaves section~\ref{sec:conclusion} for the final remarks on our work.

\section{Utility and limitation of perturbation theory of neutrino oscillation} 
\label{sec:utility-of-Ptheory}

It appears to the authors that there is no sensible consensus in the community about the values of perturbative treatment of neutrino oscillation. In this section, we try to give a balanced view on utility and limitation of such treatment, even though it is inevitably based on our prejudices. 
We do it under a hope that forming a consensus on values of perturbative treatment eventually leads to a further development of the field. 

\subsection{Utility in data analyses}
\label{sec:data-analysis}

An extremist critical statement which is often made is that there is no need for perturbation theory of neutrino oscillation because the three-flavor neutrino evolution is so simple to solve numerically even with handy personal computers nowadays. In fact, it seems that most of the numerical analyses of the experimental data are done in this way, producing a feeling that numerical computation suffices in the data analyses.\footnote{
%%%%%%%%%%%%% footnote %%%%%%%%%%%%
The situation, however, might change when we have to deal with a huge number of systematic errors and nuisance parameters in precision era of the neutrino experiments, as already hinted, e.g., in Super-Kamiokande atmospheric neutrino analysis.}
Another problem in using the analytic formulas in data analyses would be that they often rely on idealistic approximations such as uniform matter density approximation in the earth. Then, the users cannot be confident about numerical accuracies of the results. 

\subsection{Utility in qualitative understanding of neutrino oscillation}
\label{sec:qualitative}

We would like to emphasize our view that utility of perturbative formulation of neutrino oscillation exists in a completely different realm, offering the way of understanding the phenomenon of neutrino oscillation at a qualitative level.

The best example showing this feature is the understanding of parameter degeneracy offered by the approximate formulas of the oscillation probability derived by Arafune {\it et al.} \cite{Arafune:1997hd}, and Cervera {\it et al.} \cite{Cervera:2000kp}.\footnote{
%%%%%%%%%%%%%% footnote %%%%%%%%%%%%
Another examples to which we do not enter into the details is that an explicit computation of probabilities reveals new regularity such as correlation between the mixing parameters. A quite limited list includes the one between $\theta_{13}$ and the NSI parameters \cite{Kikuchi:2008vq}, and the one between $\nu$SM CP phase and the phases which represent the effects of new physics \cite{Martinez-Soler:2018lcy}.
} 
The former (the latter) applies to the case of weak (sizeable) matter effects in the LBL accelerator neutrino experiments. Using the formulas it has been recognized that there exists the three sets of parameter degeneracy, the phenomenon of multiple solutions for a given set of observed probabilities. They are, so called, the intrinsic $\theta_{13} - \delta$ degeneracy \cite{BurguetCastell:2001ez}, the sign-$\Delta m^2_{31}$ degeneracy (which implies different solutions depending upon the mass orderings) \cite{Minakata:2001qm}, and the $\theta_{23} - \delta$ degeneracy \cite{Fogli:1996pv}. Even with exact analytic expressions of the oscillation probability derived by Kimura {\it et al.} (KTY) \cite{Kimura:2002wd}, and by Zaglauer and Schwarzer (ZS) \cite{Zaglauer:1988gz}, it would have been difficult to recognize such structure of degeneracy unless these structure-transparent approximate formulas were known. 

Notice that the Cervera et al. formula can be derived by perturbative treatment to second order in the small parameters, $\epsilon \equiv \Delta m^2_{21} / \Delta m^2_{31}$ and $s_{13}$. The formula is not super-accurate, since the measured value of $\theta_{13}$ is large, $s^2_{13} = 0.022 \sim \epsilon = 0.031$ 
(see e.g., refs.~\cite{Tanabashi:2018oca,Esteban:2018azc,Capozzi:2018ubv,deSalas:2017kay}), and it requires higher order corrections to $s^4_{13}$ to match for second-order accuracy in $\epsilon$ \cite{Asano:2011nj}. Nonetheless, it did not affect utility of the formula to unfold the structure of parameter degeneracy. The experience tells us that numerical accuracy of the perturbative formula is not the whole story. 

\subsection{Utility in having framework of systematic inclusion of sub-leading effects}
\label{sec:systematic-subleading}

Another point for utility of perturbative formulation of the standard three-flavor neutrino oscillation is that it offers a systematic way of including the sub-leading effects beyond the $\nu$SM. 

One might argue that since such effect is small in size one can perturb with the small sub-leading effect itself. However, it is not practical even for constant matter density, given the complexity of the KTY or ZS solutions as the leading term. Therefore, one need a suitable perturbative framework of three neutrino evolution to include beyond the $\nu$SM sub-leading effect. 
Examples of the effects under discussion include non-standard interactions (NSI) \cite{Ohlsson:2012kf,Miranda:2015dra}, and non-unitarily due to possible new physics at high- or low-energy scales \cite{Antusch:2006vwa,Escrihuela:2015wra,Fong:2016yyh,Fong:2017gke}. Perturbative treatments with such effect were tried, e.g., in \cite{Kikuchi:2008vq} and \cite{Martinez-Soler:2018lcy} for NSI and non-unitarily, respectively. We note that having formalism for including such sub-leading effects {\em systematically} may become more important as accuracy of measurement further improves. 

\subsection{Limitation of perturbative formula}
\label{sec:limitation}

Yet, we must keep in mind that perturbative expression of the oscillation probability has many limitations. One of them, which is most important in our opinion is that the kinematical region, e.g., in energy $E$ and baseline $L$ space, of validity of the formula is quite restricted depending upon how the perturbation theory is formulated. 

An example of this feature is that perturbation theory to treat neutrino oscillation in region around the ``atmospheric-scale resonance'' fails to make proper treatment of the eigenvalue level crossing in the ``solar-scale resonance'' region \cite{Minakata:2015gra}, and vice versa.\footnote{
%%%%%%%%%%%%% footnote %%%%%%%%%%%%%%
They serve for understanding the features of neutrino oscillations in LBL accelerator or atmospheric neutrino experiments with detector at around the atmospheric oscillation maximum $\Delta m^2_{31} L / 4E = \pi / 2$ (e.g., \cite{Minakata:2015gra}), and the solar oscillation maximum $\Delta m^2_{21} L / 4E = \pi / 2$ (this paper), respectively. 
}
However, we must emphasize that it is perfectly normal in perturbation theory, nothing wrong with it by itself. Nevertheless, it must be kept in mind that we must know the region of validity of each perturbative framework.

It may be appropriate to mention here that there is an approach to remedy the problem of perturbative treatment described above, as initiated by the authors of ref.~\cite{Agarwalla:2013tza}, who performed a 1-2 space rotation first and then a 1-3 space rotation to approximately diagonalize the Hamiltonian. It was followed by Denton {\it et al.} who did the two rotations in the opposite order \cite{Denton:2016wmg}. Here, we do not discuss this approach further, regarding it a method for approximate diagonalization of Hamiltonian rather than perturbation theory, because the leading order term itself has a sufficient accuracy for most purposes. 

\subsection{Summary}
\label{sec:summary}

To summarize our discussion in this section, with full of the prejudices, we quote the following three statements:
\begin{itemize}
\item 
A perturbative formula of neutrino oscillation probability may be useful for data analyses, if it is reasonably simple and accurate, and can handle a realistic situation such as adiabatically varying matter density. 

\item 
A perturbative formula can reveal qualitative features of neutrino oscillation as illustrated by an example of how the phenomenon of parameter degeneracy was uncovered.

\item 
Perturbation theory has an inherent limitation that the applicability region of the formula is restricted to a particular region depending on how it is formulated. It is crucially important to know the region of validity. 

\end{itemize}

\section{What is new in the solar resonance perturbation theory?}
\label{sec:what-is-new}

First of all, we want to place our perturbative framework, ``solar resonance perturbation theory'', into a context in our discussion in the previous section. It is a perturbation theory formulated to treat the solar resonance region in the scope of atmospheric neutrino observation at low energies, typically $E \lsim 500$ MeV, or possibly future accelerator LBL experiments in the similar energy region. It was partly motivated as an complement to ``atmospheric resonance perturbation theory'' \cite{Minakata:2015gra}, which perturbs around atmospheric -scale enhancement. See also refs.~\cite{Cervera:2000kp,Arafune:1996bt,Freund:2001pn,Akhmedov:2004ny} for earlier references.  

Now, what is new in our perturbation theory as a framework itself? First, we must note that this is a somewhat counterintuitive way of formulating perturbation theory. We aim at perturbing the neutrino evolution by taking the solar 1-2 sector as the non-perturbative part of the Hamiltonian. A potential issue here is that $\Delta m^2_{31}$  is larger than $\Delta m^2_{21}$ by a factor of $\sim 30$, which means that there is much bigger effect outside the 1-2 zeroth order part of the Hamiltonian. It necessitates a formulation of perturbation theory in such a way that the much larger $\Delta m_{31}$ term does not disturb the zeroth-order effect. In fact, we will do a much better job, which is done in the following two ways: In leading order in expansion the $\Delta m^2_{31}$ term decouples, and in the first-order corrections, the $\Delta m^2_{31}$ shows up only in energy denominators. Consequently, we have the effective expansion parameter $A_{ \text{exp} } = c_{13} s_{13} \left( a / \Delta m^2_{31} \right) \simeq 10^{-3}$ for $E=100$ MeV [see (\ref{expansion-parameter})], where $a$ denotes the matter potential. That is, the larger the $\Delta m_{31}$, the {\em smaller} the higher order corrections. 

Then, what is new in the outcome of our perturbative formulation in physics context? The most important feature we can clearly observe by analytic means is that CP violation effect is much larger than the one we expect to see in the conventional accelerator LBL experiments, such as T2HK and DUNE. It is larger by a factor of $\sim$10 at the level of the appearance oscillation probability, the fact best illustrated by using the bi-probability plot \cite{Minakata:2001qm} as shown in figure~\ref{fig:P-barP-biP-plot}. 
The mechanism behind such large amplification is that one of the suppression factors $\propto \Delta m^2_{21} / \Delta m^2_{31}$ on $\delta$ dependent terms is dynamically lifted, as will be shown in section~\ref{sec:CP-violation}. 

In consistent with our own suggestion we examined the region of applicability of our solar resonance perturbation theory. We identified, as a necessary condition (see section~\ref{sec:how-accurate}), the region of validity of our formula as $L \gsim \ell_{ \text{atm} } = 4\pi E / \Delta m^2_{31}$. 
The smallness of the first order correction terms that comes from the large hierarchy $\Delta m^2_{31} \gg \Delta m^2_{21}$ can be confirmed as a small corrections to the zeroth-order oscillation probability, as will be seen in figure~\ref{fig:Pmue-mumu}. 
To our surprise we have found that the oscillation probability formula works in a reasonable accuracy for $L=300$ km and $E \lsim 1$ GeV, suggesting its applicability to the T2K and T2HK experiments. 

\section{Formulating perturbation theory of neutrino oscillation around the solar resonance }
\label{sec:formulation}

The standard three-flavor neutrino evolution in matter can be described by the Schr\"odinger equation in the flavor basis, $i \frac{d}{dx} \nu = H \nu$, with Hamiltonian 
\begin{eqnarray}
H= 
\frac{ 1 }{ 2E } \left\{ 
U \left[
\begin{array}{ccc}
0 & 0 & 0 \\
0 & \Delta m^2_{21}& 0 \\
0 & 0 & \Delta m^2_{31} 
\end{array}
\right] U^{\dagger}
+
\left[
\begin{array}{ccc}
a(x) & 0 & 0 \\
0 & 0 & 0 \\
0 & 0 & 0
\end{array}
\right] 
\right\}, 
\label{hamiltonian}
\end{eqnarray}
where $E$ is neutrino energy and $\Delta m^2_{ji} \equiv m^2_{j} - m^2_{i}$. 
In (\ref{hamiltonian}), $U$ denotes the standard $3 \times 3$ MNS lepton flavor mixing matrix which relates the flavor neutrino states to the vacuum mass eigenstates as $\nu_{\alpha} = U_{\alpha i} \nu_{i}$, where $\alpha$ runs over $e, \mu, \tau$, and the mass eigenstate indix $i$ runs over $1,2,$ and $3$. With the obvious notations $s_{ij} \equiv \sin \theta_{ij}$ etc. and $\delta$ for lepton version of Kobayashi-Maskawa CP violating phase, $U$ is given by
\begin{eqnarray}
U &=& 
\left[
\begin{array}{ccc}
1 & 0 &  0  \\
0 & 1 & 0 \\
0 & 0 & e^{- i \delta} \\
\end{array}
\right] 
U_{\text{\tiny PDG}} 
\left[
\begin{array}{ccc}
1 & 0 &  0  \\
0 & 1 & 0 \\
0 & 0 & e^{ i \delta} \\
\end{array}
\right] 
\nonumber \\
&=&
\left[
\begin{array}{ccc}
1 & 0 &  0  \\
0 & c_{23} & s_{23} e^{ i \delta} \\
0 & - s_{23} e^{- i \delta} & c_{23} \\
\end{array}
\right] 
\left[
\begin{array}{ccc}
c_{13}  & 0 &  s_{13} \\
0 & 1 & 0 \\
- s_{13} & 0 & c_{13}  \\
\end{array}
\right] 
\left[
\begin{array}{ccc}
c_{12} & s_{12}  &  0  \\
- s_{12} & c_{12} & 0 \\
0 & 0 & 1 \\
\end{array}
\right] 
\equiv 
U_{23} U_{13} U_{12} 
\label{MNS-matrix}
\end{eqnarray}
where we have defined $U$ matrix in a convention used in \cite{Minakata:2015gra}, which is slightly different (but physically equivalent) from $U_{\text{\tiny PDG}}$ of Particle Data Group \cite{Tanabashi:2018oca}. 

The functions $a(x)$ in (\ref{hamiltonian}) denote the Wolfenstein's matter potential \cite{Wolfenstein:1977ue} due to charged current (CC) reactions 
\begin{eqnarray} 
a &=&  
2 \sqrt{2} G_F N_e E \approx 1.52 \times 10^{-4} \left( \frac{Y_e \rho}{\rm g\,cm^{-3}} \right) \left( \frac{E}{\rm GeV} \right) {\rm eV}^2.  
\label{matt-potential}
\end{eqnarray}
Here, $G_F$ is the Fermi constant, $N_e$ is the electron number density in matter. $\rho$ and $Y_e$ denote, respectively, the matter density and number of electron per nucleon in matter. For simplicity and clarity we will work with the uniform matter density approximation in this paper. But, it is not difficult to extend our treatment to varying matter density case if adiabaticity holds.

\subsection{Relevant kinematical region}
\label{sec:kinematical-region}

We are interested in exploring the region where neutrino energy $E=( 1 - 5 ) \times 100$ MeV and baseline $L= ( 1 - 10 ) \times 1000$ km. We note that the region is around the vacuum solar oscillation maximum, $\frac{ \Delta m^2_{21} L}{4 E} \sim \mathcal{O} (1)$:
\begin{eqnarray}
\frac{ \Delta m^2_{21} L}{4 E} 
&=&
%1.27 x 10 x 7.5 x 10^{-2}
0.953
\left(\frac{\Delta m^2_{21}}{7.5 \times 10^{-5}\mbox{eV}^2}\right)
\left(\frac{L}{1000 \mbox{km}}\right)
\left(\frac{E}{100 \mbox{MeV}}\right)^{-1}.
\label{kinematic2}
\end{eqnarray}
But, since the matter potential is comparable to the vacuum effect (represented by $\Delta m^2_{21}$) in this region, 
\begin{eqnarray} 
\frac{ a }{ \Delta m^2_{21} } 
&=& 
%%2.13 \times 10^{-3} \left(\frac{ \Delta m^2 }{ 0.1~\mbox{eV}^2}\right)^{-1} \left(\frac{\rho}{2.8 \,\text{g/cm}^3}\right) \left(\frac{E}{1~\mbox{GeV}}\right) 
%
%% 2.13 x 10^{-3} x 0.1 / (7.5 x 10^{-5} ) x 0.1 = (2.13 / 7.5) = 0.284 x (3 / 2.8) (200 / 100) 
0.609 
\left(\frac{ \Delta m^2_{21} }{ 7.5 \times 10^{-5}~\mbox{eV}^2}\right)^{-1}
\left(\frac{\rho}{3.0 \,\text{g/cm}^3}\right) \left(\frac{E}{200~\mbox{MeV}}\right) 
\sim \mathcal{O} (1), 
\label{a/Dm2solar}
\end{eqnarray}
the atmospheric neutrinos at low energies, $\sim$ a few $\times 100$ MeV, are fully affected by the earth matter effect. We will see later that the role of effective expansion parameter is played by the quantity $A_{ \text{exp} } = c_{13} s_{13} \left( a / \Delta m^2_{31} \right) \simeq 10^{-3}$ for $E=100$ MeV [see (\ref{expansion-parameter})].

\subsection{Intermediate basis, or the check basis}
\label{sec:intermediate-basis} 

To formulate the solar resonance perturbation theory we transform from the flavor basis to an intermediate basis, or the check basis 
\begin{eqnarray} 
\check{\nu}_{\alpha} = \left( U_{23} U_{13} \right)^{\dagger}_{\alpha \beta} \nu_{\beta}, 
\label{check-basis}
\end{eqnarray}
with Hamiltonian 
\begin{eqnarray} 
&& 
\check{H} = \left( U_{23} U_{13} \right)^{\dagger} H U_{23} U_{13} 
%
%\nonumber \\&=&
= U_{12} 
\left[
\begin{array}{ccc}
0 & 0 & 0 \\
0 & \Delta_{21} & 0 \\
0 & 0 & \Delta_{31} \\
\end{array}
\right] 
U_{12}^{\dagger} 
+ U_{13}^{\dagger} 
\left[
\begin{array}{ccc}
\Delta_{a} & 0 & 0 \\
0 & 0 & 0 \\
0 & 0 & 0 \\
\end{array}
\right] 
U_{13}, 
\label{check-H-def}
\end{eqnarray}
where $U_{23}$ rotation on the second term is performed with no effect. Here, we have introduced simplified notations ($i, j = 1,2,3$)
\begin{eqnarray} 
\Delta_{ji} \equiv \frac{ \Delta m^2_{ji} }{ 2E }, 
\hspace{10mm}
\Delta_{a} \equiv \frac{ a }{ 2E }. 
\label{Deltas-def}
\end{eqnarray}
Using the parametrization of $U$ matrix in (\ref{MNS-matrix}) $\check{H}$ can be written as 
\begin{eqnarray} 
&& \check{H} = 
\left[
\begin{array}{ccc}
s^2_{12} \Delta_{21} & c_{12} s_{12} \Delta_{21} & 0 \\
c_{12} s_{12} \Delta_{21} & c^2_{12} \Delta_{21} & 0 \\
0 & 0 & \Delta_{31} \\
\end{array}
\right] 
+ 
\left[
\begin{array}{ccc}
c^2_{13} \Delta_{a} & 0 & c_{13} s_{13} \Delta_{a} \\
0 & 0 & 0 \\
c_{13} s_{13} \Delta_{a} & 0 & s^2_{13} \Delta_{a} \\
\end{array}
\right].
\label{check-Hamiltonian}
\end{eqnarray}

\subsection{Formulating perturbation theory with hat basis}
\label{sec:formulating-P} 

We use the ``renormalized basis'' such that the zeroth-order and the perturbed Hamiltonian takes the form $\check{H} = \check{H}_{0} + \check{H}_{1}$: 
\begin{eqnarray} 
&& \check{H}_{0} = 
\left[
\begin{array}{ccc}
s^2_{12} \Delta_{21} + c^2_{13} \Delta_{a} & c_{12} s_{12} \Delta_{21} & 0 \\
c_{12} s_{12} \Delta_{21} & c^2_{12} \Delta_{21} & 0 \\
0 & 0 & \Delta_{31} + s^2_{13} \Delta_{a}  \\
\end{array}
\right], 
\nonumber \\
&& 
\check{H}_{1} = 
\left[
\begin{array}{ccc}
0 & 0 & c_{13} s_{13} \Delta_{a} \\
0 & 0 & 0 \\
c_{13} s_{13} \Delta_{a} & 0 & 0 \\
\end{array}
\right]. 
\label{check-H-zeroth-1st}
\end{eqnarray}
We then transform to the ``hat basis'' 
\begin{eqnarray} 
\hat{\nu}_{\alpha}= ( U_{\varphi}^{\dagger} )_{\alpha \beta} \check{\nu}_{\beta}, 
\label{hat-basis}
\end{eqnarray}
with Hamiltonian 
\begin{eqnarray} 
\hat{H} = U_{\varphi}^{\dagger} \check{H} U_{\varphi} 
\label{H-hat-basis}
\end{eqnarray}
where $U_{\varphi}$ is parametrized as 
\begin{eqnarray} 
U_{\varphi} = 
\left[
\begin{array}{ccc}
\cos \varphi & \sin \varphi & 0 \\
- \sin \varphi & \cos \varphi & 0 \\
0 & 0 & 1 \\
\end{array}
\right]. 
\label{U-varphi-def}
\end{eqnarray}
$U_{\varphi}$ is determined such that $\hat{H}$ is diagonal. The condition reads 
\begin{eqnarray} 
\cos 2 \varphi &=& 
\frac{ \cos 2\theta_{12} - c^2_{13} r_{a} }
{ \sqrt{ \left( \cos 2\theta_{12} - c^2_{13} r_{a} \right)^2 +  \sin^2 2\theta_{12} } }, 
\nonumber \\
\sin 2 \varphi &=& 
\frac{ \sin 2\theta_{12} }
{ \sqrt{ \left( \cos 2\theta_{12} - c^2_{13} r_{a} \right)^2 +  \sin^2 2\theta_{12} } }, 
\label{cos-sin-2varphi}
\end{eqnarray}
where 
\begin{eqnarray} 
r_{a} \equiv \frac{a}{\Delta m^{2}_{21}} = \frac{ \Delta_{a} }{ \Delta_{21} }.
\label{ra-def}
\end{eqnarray}
The eigenvalues of the zeroth order Hamiltonian $\check{H}_{0}$ in (\ref{check-H-zeroth-1st}) is given by 
\begin{eqnarray} 
h_{1} &=& 
\sin^2 \left( \varphi - \theta_{12} \right) \Delta_{21} + \cos^2 \varphi c^2_{13} \Delta_{a},
\nonumber \\
h_{2} &=&
\cos^2 \left( \varphi - \theta_{12} \right) \Delta_{21} + \sin^2 \varphi c^2_{13} \Delta_{a},
\nonumber \\
h_{3} &=& 
\Delta_{31} + s^2_{13} \Delta_{a}. 
\label{eigenvalues}
\end{eqnarray}
Notice that one can show that 
\begin{eqnarray} 
h_{1} &=& 
\frac{ \Delta_{21} }{ 2 } 
\left[
\left( 1 + c^2_{13} r_{a} \right) 
- \sqrt{ \left( \cos 2\theta_{12} - c^2_{13} r_{a} \right)^2 +  \sin^2 2\theta_{12} } 
\right],
\nonumber \\
h_{2} &=&
\frac{ \Delta_{21} }{ 2 } 
\left[
\left( 1 + c^2_{13} r_{a} \right) 
+ \sqrt{ \left( \cos 2\theta_{12} - c^2_{13} r_{a} \right)^2 +  \sin^2 2\theta_{12} } 
\right].
\label{eigenvalues2}
\end{eqnarray}
Then, the Hamiltonian in the hat basis is given by $\hat{H} = \hat{H}_{0} + \hat{H}_{1}$ where 
\begin{eqnarray} 
\hat{H}_{0} &=& 
\left[
\begin{array}{ccc}
h_{1} & 0 & 0 \\
0 & h_{2} & 0 \\
0 & 0 & h_{3} \\
\end{array}
\right], 
\hspace{10mm}
%\nonumber \\
\hat{H}_{1} = 
\left[
\begin{array}{ccc}
0 & 0 & c_{\varphi} c_{13} s_{13} \Delta_{a} \\
0 & 0 & s_{\varphi} c_{13} s_{13} \Delta_{a} \\
c_{\varphi} c_{13} s_{13} \Delta_{a} & s_{\varphi} c_{13} s_{13} \Delta_{a} & 0 \\
\end{array}
\right],  
\label{hat-H-0th-1st}
\end{eqnarray}
where we should note that $\hat{H}_{1} = \check{H}_{1}$.

If we treat anti-neutrino channel, the transformations $\delta \rightarrow - \delta$ and $a \rightarrow -a$ suffice. We denote the angle $\varphi$ and the eigenvalues $h_{i}$ transformed by the above transformations as $\bar{\varphi}$ and $\bar{h}_{i}$. Notice that our framework is constructed such as to allow both the normal ($\Delta m^2_{31} > 0$) and the inverted ($\Delta m^2_{31} < 0$) mass orderings. Moreover, $h_{1}$, $h_{2}$ and $\varphi$ do not depend on the mass orderings because $\Delta m^2_{21} > 0$, whereas $h_{3}$ does. Therefore, unlike the case of helio-to-terrestrial ratio ($\epsilon \equiv \Delta m^2_{21} / \Delta m^2_{31}$) perturbation theory, no essential classification of the mass ordering is necessary in our present formalism.  

\subsection{Calculation of $\hat{S}$ matrix}
\label{sec:calculating-hatS}

We formulate perturbation theory in the ``hat'' basis. The $\hat{S}$ matrix is given by 
\begin{eqnarray} 
\hat{S} = T \exp \left[ -i \int^{x'}_{0} dx \hat{H} (x') \right] 
= e^{ - i \hat{H} (x) x }
\label{hatS-matrix}
\end{eqnarray}
where $T$ symbol indicates the ``space ordering'', but it simplifies for uniform matter density as shown in the second equality in eq.~(\ref{hatS-matrix}).

To calculate $\hat {S} (x)$ we define $\Omega(x)$ as
\begin{eqnarray} 
\Omega(x) = e^{i \hat{H}_{0} x} \hat{S} (x).
\label{def-omega}
\end{eqnarray}
Using $i \frac{d}{dx} \hat{S} = \hat{H} (x) \hat{S}$, $\Omega(x)$ obeys the evolution equation
\begin{eqnarray} 
i \frac{d}{dx} \Omega(x) = H_{1} \Omega(x) 
\label{omega-evolution}
\end{eqnarray}
where
\begin{eqnarray} 
H_{1} \equiv e^{i \hat{H}_{0} x} \hat{H}_{1} e^{-i \hat{H}_{0} x} .
\label{def-H1}
\end{eqnarray}
Then, $\Omega(x)$ can be computed perturbatively as
\begin{eqnarray} 
\Omega(x) &=& 1 + 
(-i) \int^{x}_{0} dx' H_{1} (x') + 
(-i)^2 \int^{x}_{0} dx' H_{1} (x') \int^{x'}_{0} dx'' H_{1} (x'') 
+ \cdot \cdot \cdot,
\label{Omega-expansion}
\end{eqnarray}
and the $\hat{S}$ matrix is given by
\begin{eqnarray} 
\hat{S} (x) =  
e^{-i \hat{H}_{0} x} \Omega(x). 
\label{hat-Smatrix}
\end{eqnarray}
Noticing that 
\begin{eqnarray} 
e^{ \pm i \hat{H}_{0} x } &=& 
\left[
\begin{array}{ccc}
e^{ \pm i h_{1} x } & 0 & 0 \\
0 & e^{ \pm i h_{2} x } & 0 \\
0 & 0 & e^{ \pm i h_{3} x } \\
\end{array}
\right], 
\end{eqnarray}
$H_{1}$ is given by 
\begin{eqnarray} 
H_{1} &=& c_{13} s_{13} \Delta_{a} 
\left[
\begin{array}{ccc}
0 & 0 & c_{\varphi} e^{ - i ( h_{3} - h_{1} ) x } \\
0 & 0 & s_{\varphi} e^{ - i ( h_{3} - h_{2} ) x } \\
c_{\varphi} e^{ i ( h_{3} - h_{1} ) x } & s_{\varphi} e^{ i ( h_{3} - h_{2} ) x } & 0 \\
\end{array}
\right].  
\label{H1}
\end{eqnarray}
$\Omega(x)$ can be calculated by using (\ref{Omega-expansion}) with (\ref{H1}). By using (\ref{hat-Smatrix}), $\hat{S}$ matrix is given to first order in $H_{1}$ by 
\begin{eqnarray} 
&& \hat{S} (x) =  
e^{-i \hat{H}_{0} x} \Omega(x) 
\nonumber \\
&& \hspace{-14mm} 
= 
\left[
\begin{array}{ccc}
e^{ - i h_{1} x } & 0 & c_{\varphi} c_{13} s_{13} \frac{ \Delta_{a} }{ h_{3} - h_{1} } 
\left\{ e^{ - i h_{3} x } - e^{ - i h_{1} x }  \right\} \\
0 & e^{ - i h_{2} x } & s_{\varphi} c_{13} s_{13} \frac{ \Delta_{a} }{ h_{3} - h_{2} } 
\left\{ e^{ - i h_{3} x } - e^{ - i h_{2} x } \right\} \\
c_{\varphi} c_{13} s_{13} \frac{ \Delta_{a} }{ h_{3} - h_{1} } 
\left\{ e^{ - i h_{3} x } - e^{ - i h_{1} x } \right\} & s_{\varphi} c_{13} s_{13} \frac{ \Delta_{a} }{ h_{3} - h_{2} } 
\left\{ e^{ - i h_{3} x } - e^{ - i h_{2} x } \right\} & e^{ - i h_{3} x } \\
\end{array}
\right]. 
\nonumber \\
\label{hat-S-matrix-1st}
\end{eqnarray}
Notice that the large element outside ``solar'' $1-2$ sector of $\check{H}_{0}$ appears solely in the energy denominator.\footnote{
%%%%%%%%%%%%%% 
The smallness of the first order correction is noticed in a different language in ref.~\cite{Peres:2003wd}. 
}
Therefore, as we mentioned earlier, the effective expansion parameter becomes 
\begin{eqnarray}
A_{ \text{exp} } 
&\equiv& 
c_{13} s_{13} 
\biggl | \frac{ a }{ \Delta m^2_{31} } \biggr | %%expansion-parameter
%\nonumber \\&=&
%0.8875 \times 10^{-2} x 0.146 = 0.1296 \times 10^{-2} x (3/2.8) 2 (200MeV) = 0.278 \times 10^{-2} 
= 2.78 \times 10^{-3} 
\left(\frac{ \Delta m^2_{31} }{ 2.4 \times 10^{-3}~\mbox{eV}^2}\right)^{-1}
\left(\frac{\rho}{3.0 \,\text{g/cm}^3}\right) \left(\frac{E}{200~\mbox{MeV}}\right), 
\nonumber \\
\label{expansion-parameter}
\end{eqnarray}
which is very small. It gives us a feeling that the leading term in the expansion must give a fair approximation, which will be confirmed in section~\ref{sec:how-accurate}. 
Notice also that  $\hat{S}$ respects T invariance, $\hat{S}_{ij} = \hat{S}_{ji}$, as it should. 

\subsection{Calculation of $S$ matrix and the oscillation probability}
\label{sec:calculating-S-matrix}

Since the relationship between the flavor basis and the hat basis is given by  
\begin{eqnarray}
\nu_{\alpha} &=&
\left( U_{23} U_{13} \right)_{\alpha \beta} \check{\nu}_{\beta} 
= \left( U_{23} U_{13} U_{\varphi} \right)_{\alpha \beta} \hat{\nu}_{\beta}, 
\nonumber \\
H &=& 
\left( U_{23} U_{13} \right) \check{H} \left( U_{23} U_{13} \right)^{\dagger} 
= 
\left( U_{23} U_{13} U_{\varphi} \right) \hat{H} \left( U_{23} U_{13} U_{\varphi} \right)^{\dagger}, 
\label{flavor-hat-basis}
\end{eqnarray}
the $S$ matrix in flavor basis can be readily calculated as  
\begin{eqnarray}
S &=&
\left( U_{23} U_{13} U_{\varphi} \right) \hat{S} \left( U_{23} U_{13} U_{\varphi} \right)^{\dagger}. 
\label{flavor-basis-S-matrix}
\end{eqnarray}
In appendix~\ref{sec:S-matrix}, we briefly describe the computation and give the resulting expressions of $S$ matrix elements to first order in expansion.

Then, the oscillation probability $P(\nu_{\beta} \rightarrow \nu_{\alpha})$ at baseline $x$ is given by 
\begin{eqnarray}
P(\nu_{\beta} \rightarrow \nu_{\alpha}: x) = \vert S_{\alpha \beta} (x) \vert^2. 
\label{oscillation-probability}
\end{eqnarray}
Give the expressions of $S$ matrix elements, it is straightforward to compute the oscillation probability to first order in expansion. All the expressions of the oscillation probability except for the ones which require $\nu_{\tau}$ beam are given either in the next section~\ref{sec:oscillation-probability-symmetry}, or in appendix~\ref{sec:oscillation-probability}.\footnote{
%%%%%%%%%%%%% footnote %%%%%%%%%%%%%%
If necessary, $P(\nu_{\tau} \rightarrow \nu_{e})$ and $P(\nu_{\tau} \rightarrow \nu_{\mu})$ can be obtained as T conjugate of $P(\nu_{e} \rightarrow \nu_{\tau})$ and $P(\nu_{\mu} \rightarrow \nu_{\tau})$, respectively. Then, $P(\nu_{\tau} \rightarrow \nu_{\tau})$ can be calculated by using unitarity $P(\nu_{\mu} \rightarrow \nu_{\tau}) = 1 - P(\nu_{\tau} \rightarrow \nu_{e}) - P(\nu_{\tau} \rightarrow \nu_{\mu})$. 
}

\section{The oscillation probability and its symmetry}
\label{sec:oscillation-probability-symmetry}

In this section we present the explicit form of the zeroth-order oscillation probability in the $\nu_{\mu} \rightarrow \nu_{e}$ and $\nu_{\mu} \rightarrow \nu_{\mu}$ channels. To simplify the expression we define the reduced Jarlskog factor in matter 
\begin{eqnarray} 
J_{mr} &\equiv& 
c_{23} s_{23} c^2_{13} s_{13} c_{\varphi} s_{\varphi} 
=
J_r \left[ 
\left( \cos 2\theta_{12} - c^2_{13} r_{a} \right)^2 +  \sin^2 2\theta_{12} 
\right]^{ - 1/2 }, 
\label{Jmr-def}
\end{eqnarray}
which is proportional to the reduced Jarlskog factor in vacuum, $J_r \equiv c_{23} s_{23} c^2_{13} s_{13} c_{12} s_{12}$. We have used eq.~(\ref{cos-sin-2varphi}) in the second equality in (\ref{Jmr-def}). Notice that not only $\sin \delta$ but also $\cos \delta$ terms in the oscillation probability in the $\nu_{e}$-related sector ($\nu_{\mu} - \nu_{\tau}$ sector) must be proportional to $J_r$ ($J_r / c^2_{13}$), as proved in \cite{Asano:2011nj}, and our expressions are consistent with this result.
We use the notation $x$ as baseline distance in this section, as well as in appendices~\ref{sec:S-matrix} and \ref{sec:oscillation-probability}. 

\subsection{The oscillation probability: $\nu_{\mu} \rightarrow \nu_{e}$ and $\nu_{\mu} \rightarrow \nu_{\mu}$ channels}
\label{sec:probability-mue-mumu}

$P(\nu_{\mu} \rightarrow \nu_{e})^{(0)}$ is given by T conjugate (by the transformation $\delta \rightarrow - \delta$)
of $P(\nu_{e} \rightarrow \nu_{\mu})^{(0)}$ given in (\ref{P-emu-0th}): 
\begin{eqnarray} 
&& 
P(\nu_{\mu} \rightarrow \nu_{e})^{(0)} = 
c^2_{23} c^2_{13} \sin^2 2 \varphi 
\sin^2 \frac{ ( h_{2} - h_{1} ) x }{2}
\nonumber \\
&+& 
s^2_{23} \sin^2 2\theta_{13} 
\left[
c^2_{\varphi} \sin^2 \frac{ ( h_{3} - h_{1} ) x }{2} 
+ s^2_{\varphi} \sin^2 \frac{ ( h_{3} - h_{2} ) x }{2} 
- c^2_{\varphi} s^2_{\varphi} \sin^2 \frac{ ( h_{2} - h_{1} ) x }{2} 
\right]
\nonumber \\
&+& 
4 J_{mr} \cos \delta 
\left\{
\cos 2 \varphi \sin^2 \frac{ ( h_{2} - h_{1} ) x }{2} 
- \sin^2 \frac{ ( h_{3} - h_{2} ) x }{2} 
+ \sin^2 \frac{ ( h_{3} - h_{1} ) x }{2} 
\right\} 
\nonumber \\
&+& 
8 J_{mr} \sin \delta 
\sin \frac{ ( h_{3} - h_{2} ) x }{2} 
\sin \frac{ ( h_{2} - h_{1} ) x }{2} 
\sin \frac{ ( h_{1} - h_{3} ) x }{2}. 
\label{P-mue-0th}
\end{eqnarray}
Whereas $P(\nu_{\mu} \rightarrow \nu_{\mu})^{(0)}$ takes the form 
\begin{eqnarray} 
&& 
P(\nu_{\mu} \rightarrow \nu_{\mu})^{(0)} 
%\nonumber \\&=& 
= 1 - \biggl[
4 c^2_{23} s^2_{23} s^2_{13} \cos^2 2 \varphi 
+ 
\sin^2 2\varphi
\left( c^4_{23} + s^4_{23} s^4_{13} \right)
\biggr]
\sin^2 \frac{ ( h_{2} - h_{1} ) x }{2} 
\nonumber \\
&-& 
4 s^2_{23} c^2_{13} 
\left[ 
\left( c^2_{23} s^2_{\varphi} + s^2_{23} s^2_{13} c^2_{\varphi} \right) 
\sin^2 \frac{ ( h_{3} - h_{1} ) x }{2} 
+ 
\left( c^2_{23} c^2_{\varphi} + s^2_{23} s^2_{13} s^2_{\varphi} \right) 
\sin^2 \frac{ ( h_{3} - h_{2} ) x }{2} 
\right]
\nonumber \\
&+&
8 c^2_{23} s^2_{23} s^2_{13} c^2_{\varphi} s^2_{\varphi} \cos 2\delta 
\sin^2 \frac{ ( h_{2} - h_{1} ) x }{2} 
\nonumber \\
&-&
8 c_{23} s_{23} s_{13} c_{\varphi} s_{\varphi} \cos \delta 
\nonumber \\
&\times&
\left[
\left( c^2_{23} - s^2_{23} s^2_{13} \right) \cos 2\varphi 
\sin^2 \frac{ ( h_{2} - h_{1} ) x }{2} 
+
s^2_{23} c^2_{13} 
\left\{
\sin^2 \frac{ ( h_{3} - h_{1} ) x }{2} - \sin^2 \frac{ ( h_{3} - h_{2} ) x }{2} 
\right\}
\right]. 
\nonumber \\
\label{P-mumu-0th}
\end{eqnarray}

What is a characteristic difference between the zeroth-order oscillation probability $P(\nu_{\mu} \rightarrow \nu_{e})^{(0)}$ and the one around the atmospheric oscillation maximum? For the latter, if necessary, we refer the expression in ref.~\cite{Minakata:2015gra} for definiteness. There are at least two important differences: 
\begin{itemize}

\item 
The main term of the oscillation probability is proportional to $c^2_{23}$ in our solar-resonance perturbation theory [see the first term in (\ref{P-mue-0th})], while it is proportional to $s^2_{23}$ in atmospheric-resonance perturbation theory. This property will play crucial role in resolving the $\theta_{23} - \delta$ degeneracy. A brief discussion of the $\theta_{23} - \delta$ degeneracy with our formula is given in appendix~\ref{sec:parameter-degeneracy}. 

\item 
In our solar-resonance perturbation theory the relevant dynamical variables at the solar-scale enhancement, the eigenvalues $h_{1}$, $h_{2}$, and $\varphi$, depend on $\Delta m^2_{21}$ but not on $\Delta m^2_{31}$ or $\Delta m^2_{32}$. It means that there is no essential difference in important features between the normal and the inverted mass orderings. It will simplify the treatment of parameter degeneracy because the sign-$\Delta m^2_{31}$ ambiguity essentially decouples.  

\end{itemize}

\subsection{Symmetry of the oscillation probability}
\label{sec:symmetry}

One recognizes that a symmetry exists in eqs.~(\ref{P-mue-0th}) and (\ref{P-mumu-0th}). That is, $P(\nu_{\mu} \rightarrow \nu_{e})^{(0)}$ and $P(\nu_{\mu} \rightarrow \nu_{\mu})^{(0)}$ are both invariant under the transformation 
\begin{eqnarray} 
&& 
h_{1} \rightarrow h_{2}, 
\hspace{10mm}
h_{2} \rightarrow h_{1}, 
\nonumber \\
&&
\cos 2\varphi \rightarrow - \cos 2\varphi, 
\hspace{10mm}
\sin 2\varphi \rightarrow - \sin 2\varphi.
\label{varphi-transformation}
\end{eqnarray}
In fact, it is easy to confirm that the symmetry exists in all the oscillation channels. The invariance can be understood as the one under the transformation $s_{\varphi} \rightarrow + c_{\varphi} $, $c_{\varphi} \rightarrow - s_{\varphi}$, or at the $\varphi$ level under 
\begin{eqnarray} 
&& 
\varphi \rightarrow \varphi + \frac{\pi}{2}, 
\label{varphi-transformation-summary}
\end{eqnarray}
which produces the eigenvalue exchange $h_{1} \leftrightarrow h_{2}$ owing to the relations  
$\sin (\varphi - \theta_{12}) \rightarrow \cos ( \varphi - \theta_{12} )$ and 
$\cos (\varphi - \theta_{12}) \rightarrow - \sin ( \varphi - \theta_{12} )$ under the transformation (\ref{varphi-transformation-summary}). 
It is interesting to note that the symmetry involves $\varphi$, the matter-affected mixing angle $\theta_{12}$ which descries the 1-2 level crossing in matter. 

Since the transformation (\ref{varphi-transformation-summary}) describes a reparametrization of $\varphi$, nature of the symmetry can be understood as a ``dynamical symmery'', not a symmetry of Hamiltonian. Nonetheless, recognizing it is useful for a consistency check of the calculation. 

In fact, one can show that a similar symmetry exists in the expression of the oscillation probability in ``atmospheric resonance'' perturbation theory formulated in \cite{Minakata:2015gra}. See appendix~\ref{sec:symmetry-MP-formula} for details.

\subsection{CP violation around the solar resonance}
\label{sec:CP-violation}

The most important message in this paper is that CP violation effect is large in the solar oscillation enhanced region, say $E \simeq \mbox{a few} \times 100$ MeV and $L \simeq \mbox{a few} \times 1000$ km, for which our perturbative framework serves. Roughly speaking, it is {\em $\sim$10 times larger} than the CP odd term in the oscillation probability estimated for the ongoing and the upcoming LBL experiments \cite{Abe:2018wpn,NOvA:2018gge,Abe:2015zbg,Acciarri:2015uup}.

To show the point, let us first look at the CP-odd $\sin \delta$ term in $P(\nu_\mu \rightarrow \nu_e)$ in vacuum. It takes the form as 
\begin{eqnarray}
&& - 8 J_r \sin \delta 
\sin \left( \frac{\Delta_{21} x}{ 2 } \right) 
\sin \left( \frac{\Delta_{31} x}{ 2 } \right) 
\sin \left( \frac{\Delta_{32} x}{ 2 } \right) 
\nonumber \\
&\approx&
- 8 J_r \sin \delta 
\sin \left( \frac{\Delta_{21} x}{ 2 } \right) 
\sin^2 \left( \frac{\Delta_{31} x}{ 2 } \right), 
\label{Pmue-vacuum-CPV}
\end{eqnarray}
where $\Delta_{21} \equiv \Delta m^2_{21} / (2 E)$ etc., and we have used in the second line $\epsilon \equiv \frac{ \Delta_{21} }{ \Delta_{31} } \approx \frac{ \Delta m^2_{21} }{ \Delta m^2_{31} } \simeq 1/30 \ll 1$ so that $\Delta_{32} \approx \Delta_{31}$.

Now, at around the atmospheric oscillation maximum, $\frac{\Delta_{31} x}{ 2 } \simeq 1$, the size of $\sin \delta$ term is given by 
$8 J_r \frac{\Delta_{21} x}{ 2 } \simeq 8 J_r / 30$. 
Whereas at around the solar oscillation maximum, $\sin \frac{\Delta_{21} x}{ 2 } \simeq 1$, the size of $\sin \delta$ term is given by 
$8 J_r \langle \sin^2 \left( \frac{\Delta_{31} x}{ 2 } \right) \rangle \simeq 4 J_r$, where we take average over the fast atmospheric scale oscillations. Therefore, the magnitude of CP odd term at around the solar oscillation maximum is 15 times larger than that at the atmospheric oscillation maximum. It is the latter quantity that is planned to be measured in the future LBL experiments. 
Therefore, in a nutshell, the reason why the effect of CP phase $\delta$ is large at around the solar-scale enhanced region is that the factor $\sin \frac{\Delta_{21} x}{ 2 }$, which would become a suppression factor $\frac{\Delta_{21} x}{ 2 } \simeq \Delta m^2_{21} / \Delta m^2_{31} = \epsilon$ at around the atmospheric oscillation maximum $\frac{\Delta_{31} x}{ 2 } \simeq 1$, no more acts as a suppression factor, $\sin \frac{\Delta_{21} x}{ 2 } \simeq 1$. That is, the suppression of the CP-odd term due to small $\epsilon$ is dynamically lifted by the solar-scale enhancement.

In practice, such neutrino beam passes through inside the earth, for which the matter effect must be taken into account. Here, we use the zeroth-order oscillation probability $P(\nu_{\mu} \rightarrow \nu_{e})^{(0)}$ in (\ref{P-mue-0th}) to give a rough estimation of size of the $\delta$ dependent term. 
To illuminate the point, we take average over the atmospheric-scale short wavelength oscillations.\footnote{
%%%%%%%%%%%% footnote %%%%%%%%%%%%%
We use 
\begin{eqnarray} 
&& 
\left\langle \sin^2 \frac{ ( h_{3} - h_{i} ) x }{2} \right\rangle 
\approx \frac{1}{2} ~~~~(i=1,2),
\nonumber \\
&& 
\left\langle \sin \frac{ ( h_{3} - h_{2} ) x }{2} 
\sin \frac{ ( h_{2} - h_{1} ) x }{2} 
\sin \frac{ ( h_{1} - h_{3} ) x }{2}  \right\rangle 
\approx 
- \frac{1}{4} \sin ( h_{2} - h_{1} ) x. 
\label{averaging}
\end{eqnarray}
}
Then, we obtain for $P(\nu_{\mu} \rightarrow \nu_{e})^{(0)}$
\begin{eqnarray} 
\left\langle P(\nu_{\mu} \rightarrow \nu_{e})^{(0)} \right\rangle 
&=& 
\frac{1}{2} s^2_{23} \sin^2 2\theta_{13} 
+ c^2_{13} \sin^2 2 \varphi 
\left( 
c^2_{23} - s^2_{23} s^2_{13} 
\right) 
\sin^2 \frac{ ( h_{2} - h_{1} ) x }{2} 
\nonumber \\
&+& 
4 J_{mr} \cos \delta 
\cos 2 \varphi \sin^2 \frac{ ( h_{2} - h_{1} ) x }{2} 
- 2 J_{mr} \sin \delta \sin ( h_{2} - h_{1} ) x.
\label{P-mue-0th-averaged}
\end{eqnarray}
This is essentially the two-flavor oscillation probability in matter near the resonance.

We roughly estimate the value of the $\sin \delta$ term around the solar resonance. For simplicity, let us tune the energy and baseline such that it is on resonance, $\cos 2\theta_{12} - c^2_{13} r_{a} =0$, or $\varphi = \frac{\pi}{4}$, and $\sin ( h_{2} - h_{1} ) x$ is maximal, $( h_{2} - h_{1} ) x = \pi/2$. 
%
%estimate the sizes of $\sin \delta$ and $\cos \delta$ terms at the solar resonance point, $\cos 2\theta_{12} - c^2_{13} r_{a} =0$, or $\varphi = \frac{\pi}{4}$. Let us estimate the largest value that $\sin \delta$ term can have. 
%
We use the best fit values of the mixing parameters given in \cite{Esteban:2018azc} for the normal mass ordering, which leads to 
$\sin 2\theta_{12} =0.925$   %%$=0.9248$, 
$J_r \equiv c_{23} s_{23} c^2_{13} s_{13} c_{12} s_{12} = 0.0334$, %%$= 0.03339$, 
and 
$J_{mr} = J_r / \sin 2\theta_{12} = 0.0361$.  %%$= 0.03611$. 
Then, the magnitude of the $\sin \delta$ term at the solar resonance is given by 
\begin{eqnarray} 
&& 2 J_{mr} = 2 J_r / \sin 2\theta_{12} = 0.0722  %%0.07222
\label{sin-delta-term}
\end{eqnarray}
while the $\cos \delta$ term vanishes. 
Apart from $\cos 2\varphi$ which vanishes at this particular point, the solar resonance point $\varphi = \frac{\pi}{4}$, the magnitude of the $\cos \delta$ term is also $2 J_{mr}$, the same as the $\sin \delta$ term given in (\ref{sin-delta-term}). Therefore, apart from the particular point, generically the CP-even $\cos \delta$ term has the same order of magnitude as the CP-odd term.

We want to compare the value of CP odd term in (\ref{sin-delta-term}) to the magnitudes of the $\sin \delta$ term to be seen in the future LBL experiments such as T2K/T2HK, NO$\nu$A, and DUNE \cite{Abe:2018wpn,NOvA:2018gge,Abe:2015zbg,Acciarri:2015uup}, whose set up are close to the atmospheric oscillation maximum. We use, for a rough comparison, the value calculated above assuming vacuum oscillation, $8 \epsilon J_r \simeq 0.0089$, which gives a fair, albeit crude, estimation.\footnote{
%%%%%%%%%%%%% footnote %%%%%%%%%%%%%%
The matter effect does not affect so much the sizes of $\delta$ dependent terms. The simplest way to see this is to look into the neutrino-antineutrino bi-probability plot \cite{Minakata:2001qm}. The matter potential shifts the CP ellipse to right-down (left-up) direction in the normal (inverted) mass ordering, but keeping its size almost as it is for relatively short baseline $L \sim 1000$ km.
}
Therefore, the value of CP odd term at the solar resonance in (\ref{sin-delta-term}) is 8 times larger than the CP odd term in vacuum estimated at the atmospheric oscillation maximum. One must keep in mind that this result is obtained by using (\ref{P-mue-0th-averaged}), which is obtained by averaging over the short wavelength atmospheric-scale oscillations. Since its magnitude easily reaches to a few tenth of the averaged probability it is difficult to obtain the unique value for the amplification factor. Therefore, it is fair to say that it is $\sim10$.\footnote{
%%%%%%%%%%%%%% footnote %%%%%%%%%%%%%%
Another example of amplification of the $\delta$ dependent terms is the one at the second oscillation maximum \cite{Ishitsuka:2005qi,Baussan:2013zcy}. One may regard this amplification of a factor of $\sim$3 is due to ``imperfect growth'' of the solar-scale oscillation. 
}

\begin{figure}[h!]
\begin{center}
\vspace{2mm}
\includegraphics[width=0.48\textwidth]{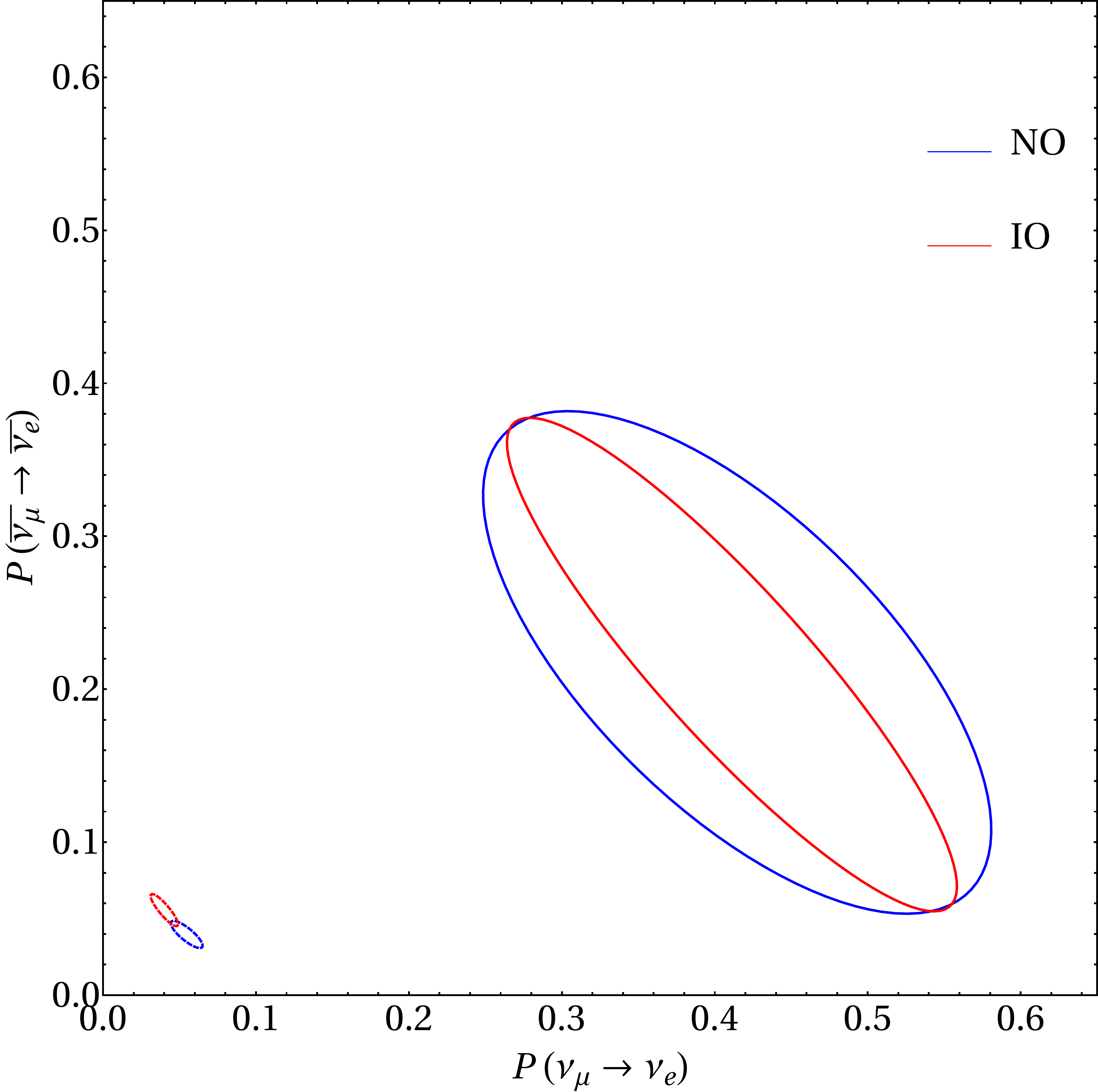}
\end{center}
\vspace{-1mm}
\caption{The bi-probability diagrams in $P(\nu_{\mu} \rightarrow \nu_{e}) - P(\bar{\nu}_{\mu} \rightarrow \bar{\nu}_{e})$ space \cite{Minakata:2001qm} are presented for the solar-scale enhanced region (large ellipses); $E=0.1$ GeV and $L=2000$ km and for NO$\nu$A setting (small ellipses); $E=2$ GeV and $L=810$ km. The blue and red lines are for the normal and inverted mass orderings, respectively.  The values of the mixing parameters are taken from ref.~\cite{Esteban:2018azc}, i.e., the best fit values for the normal mass ordering. 
} 
\vspace{-2mm}
\label{fig:P-barP-biP-plot}
\end{figure}

The contrast of the sizes of CP phase dependence between the solar-scale oscillation enhanced region and the conventional LBL setting, NO$\nu$A as just an example, can be best represented by the bi-probability plot in $P(\nu_{\mu} \rightarrow \nu_{e}) - P(\bar{\nu}_{\mu} \rightarrow \bar{\nu}_{e})$ space \cite{Minakata:2001qm}, as seen in figure~\ref{fig:P-barP-biP-plot}. In this example the ellipses in the solar-scale enhanced region are about 10 times larger than NO$\nu$A's. 
Notice that the features of bi-probability ellipse in the solar-scale enhanced region, in particular their shapes, are extremely sensitive to the values of the mixing parameters, in particular $\Delta m^2_{31}$ \cite{Nunokawa-private}, due to coexistence of the atmospheric- and the solar-scale oscillations. However, we do not enter into the complexity in this paper, restricting ourselves to a simple but conservative estimation of the amplification factor of CP phase effect.

The similar discussion on the size of the CP-odd term can be easily extended to that of the $\nu_{e} \rightarrow \nu_{\mu}$ channel, because they have the same magnitude in vacuum and in matter due to T-invariance. Similarly, the same argument must go through for the $\nu_{e} \rightarrow \nu_{\tau}$ channel, because the CP-odd term has the same size as that of the $\nu_{e} \rightarrow \nu_{\mu}$ channel in vacuum and in matter due to unitarity. 

%%%%%%%%%%%%%%% FIG 1 %%%%%%%%%%%%%%%
%%%%%%%%%%%%%%% FIG 1 %%%%%%%%%%%%%%%

\section{How accurate are the oscillation probability formulas?}
\label{sec:how-accurate}

\begin{figure}[h!]
\begin{center}
\vspace{1mm}
\includegraphics[width=0.48\textwidth]{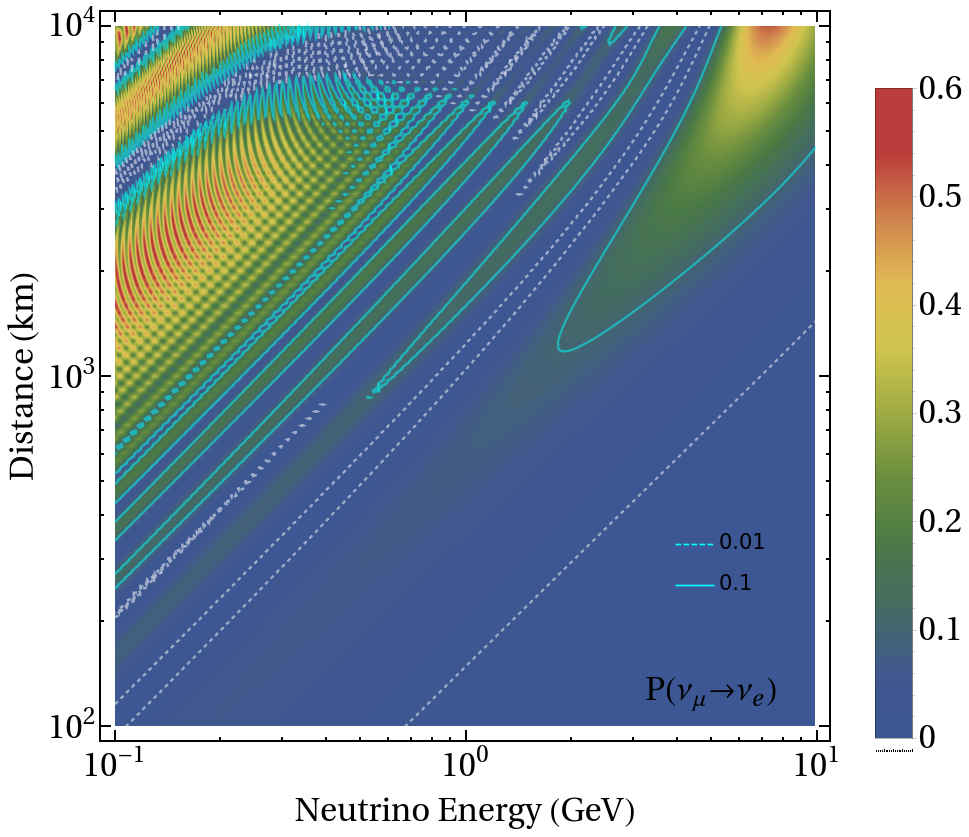}
\includegraphics[width=0.48\textwidth]{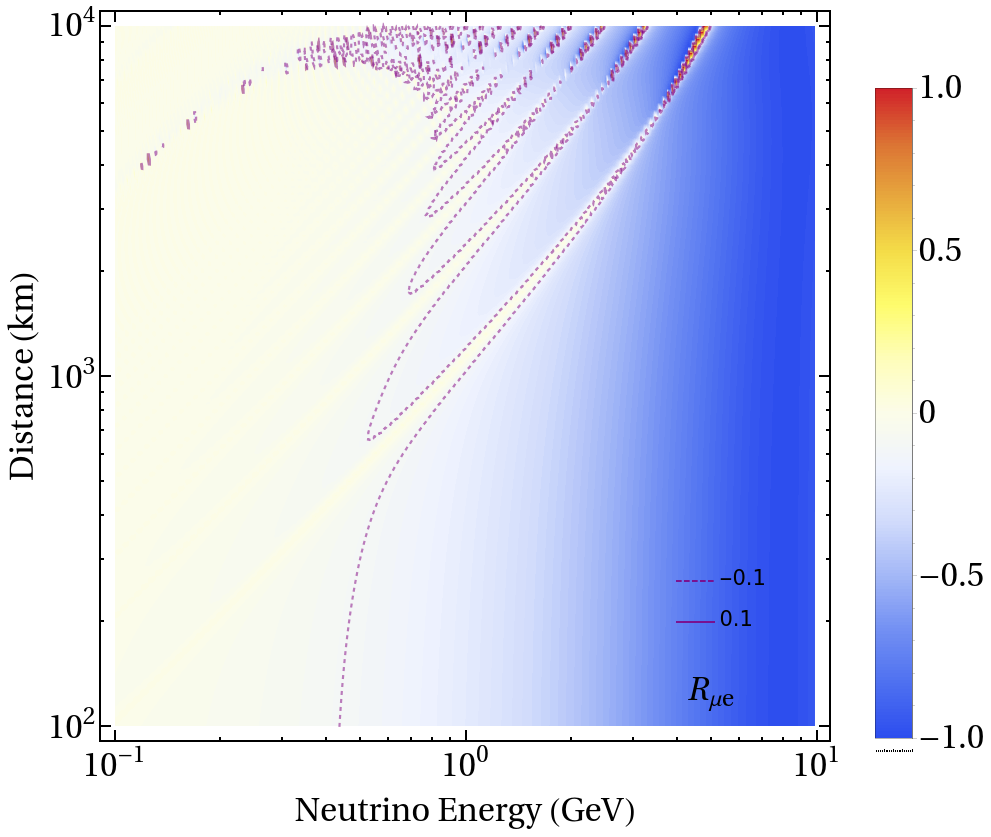}
\end{center}
\vspace{-3mm}
\caption{ In the left panel, the oscillation probability $P(\nu_{\mu} \rightarrow \nu_{e})$ is shown as the equi-probability contours superimposed on a color grade plot in neutrino energy $E$ and baseline $L$ plane. In the right panel, the ratio of deviation of the zeroth-order formula to the exact probability,  $P(\nu_{\mu} \rightarrow \nu_{e})^{(0)} / P(\nu_{\mu} \rightarrow \nu_{e})_{ \text{exact}  } -1$, is plotted in the same style. In both panels the matter density is taken to be a constant, $\rho = 4.0~\text{g/cm}^3$. The values of the mixing parameters are taken from the best fit values of ref.~\cite{Esteban:2018azc}, including $\delta = 215$ degree.
} 
\vspace{-5mm}
\label{fig:Pmue-color-grade} 
\end{figure}

\begin{figure}[h!]
\begin{center}
\vspace{-2mm}
\includegraphics[width=0.90\textwidth]{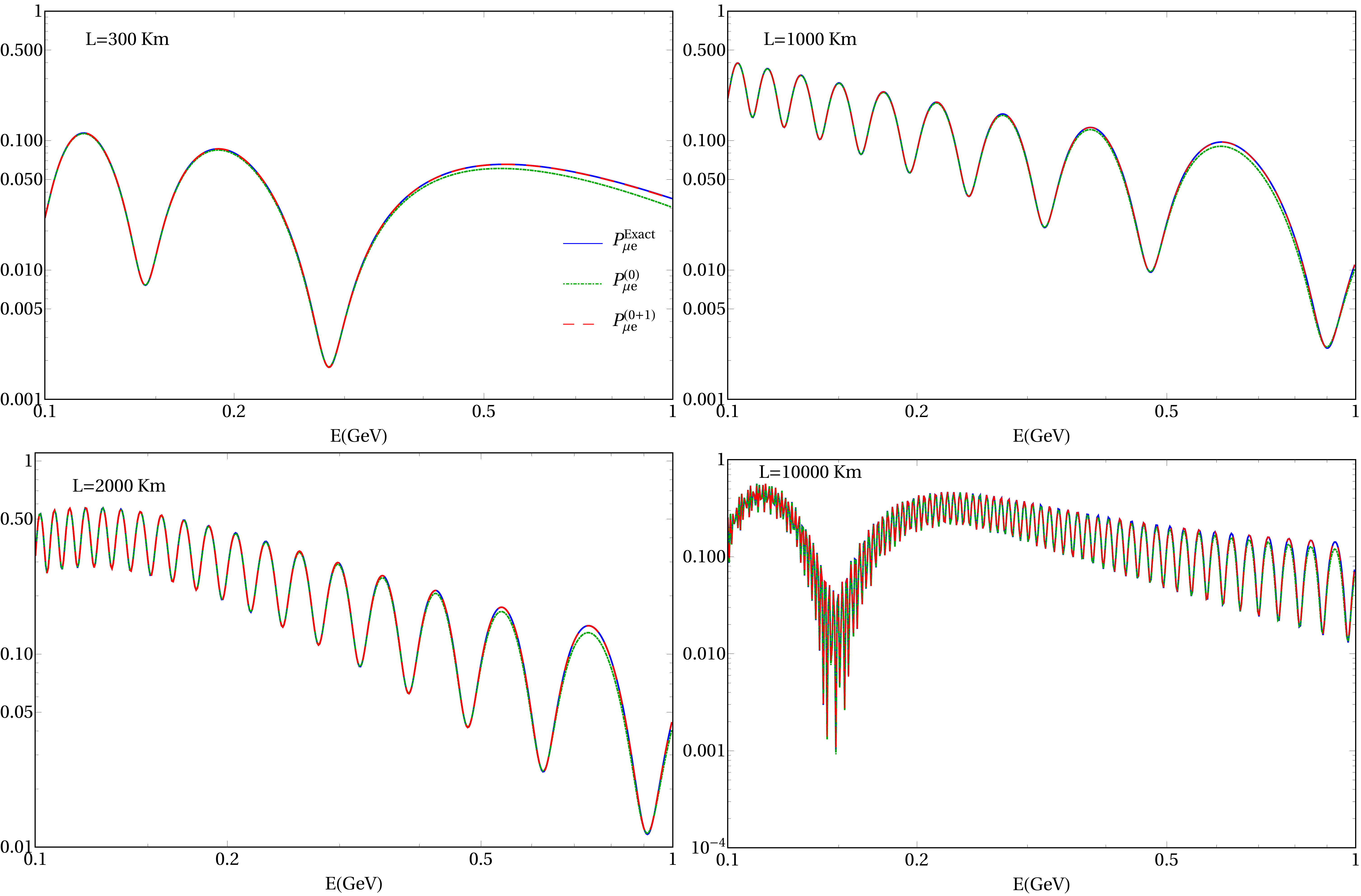}
\includegraphics[width=0.90\textwidth]{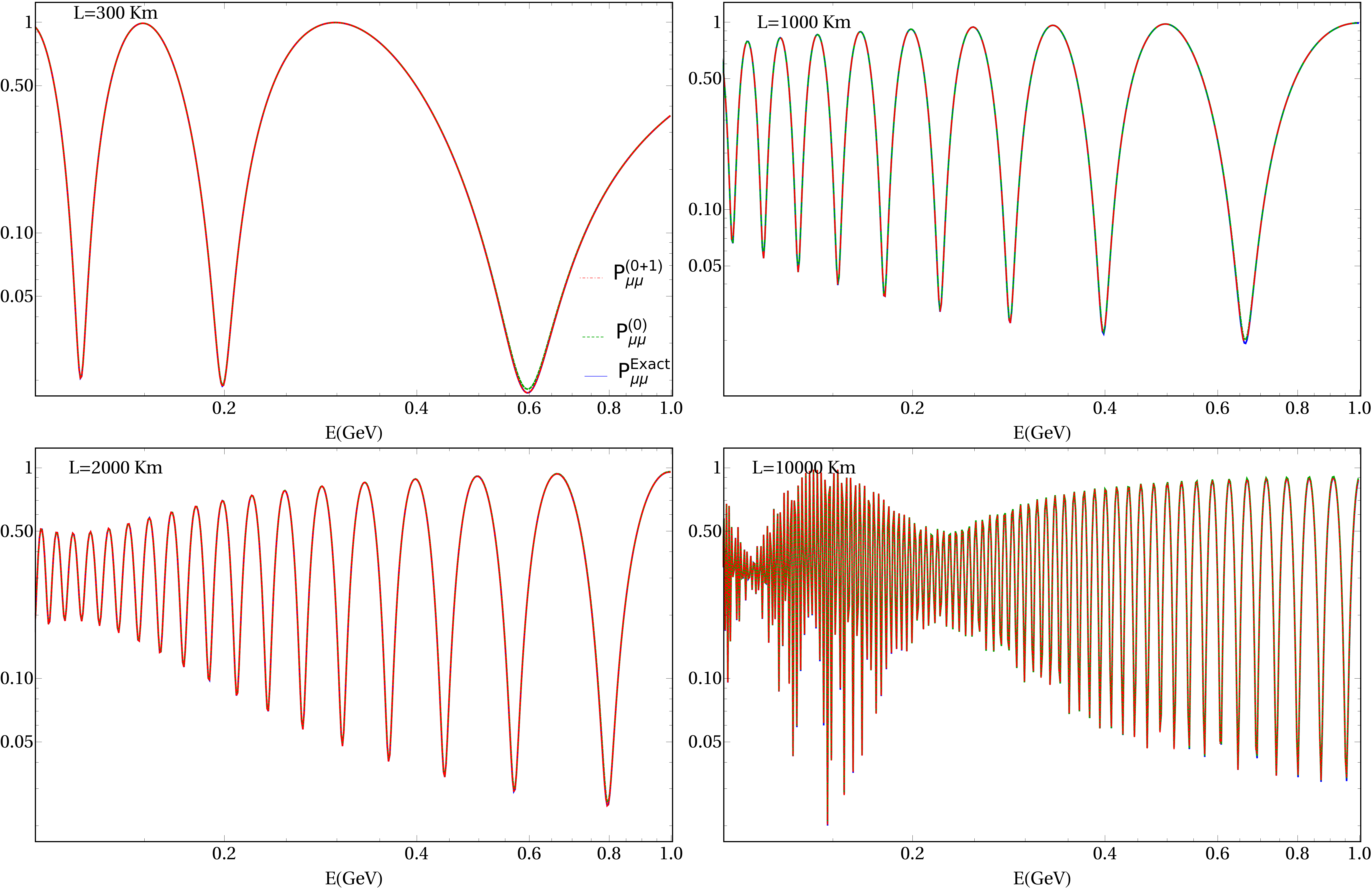}
\end{center}
\vspace{0mm}
\caption{ The energy dependences of the oscillation probabilities $P(\nu_{\mu} \rightarrow \nu_{e})$ (upper cluster of four panels), and $P(\nu_{\mu} \rightarrow \nu_{\mu})$ (lower cluster of four panels). In each cluster the four panels are for $L=300$, 1000, 2000, and 10000 km. The green dotted and the red dashed lines denote $P(\nu_{\mu} \rightarrow \nu_{\alpha})^{(0)}$ and $P(\nu_{\mu} \rightarrow \nu_{\alpha})^{(0)} + P(\nu_{\mu} \rightarrow \nu_{\alpha})^{(1)}$ ($\alpha= e, \mu$), respectively, which should be compared to the exact numerical computation given by the blue solid lines. 
}
\label{fig:Pmue-mumu} 
\end{figure}

In this section we investigate numerically the oscillation probability formulas derived by the solar resonance perturbation theory to know how accurate they are. Though the numerical accuracy is not our primary concern, as we discussed in section~\ref{sec:utility-of-Ptheory}, it is better to check how good is our expectation to the accuracy we spelled out in section~\ref{sec:formulation}.
We examine first overall (dis-) agreement between our leading order formula and the exact expression of $P(\nu_{\mu} \rightarrow \nu_{e})$, which is presented in figure~\ref{fig:Pmue-color-grade}. 
Then, in figure~\ref{fig:Pmue-mumu}, we display accuracies of the formulas in more details by comparing our leading order and the leading plus first-order formulas to the exact energy dependences of $P(\nu_{\mu} \rightarrow \nu_{e})$ and $P(\nu_{\mu} \rightarrow \nu_{\mu})$ at several baseline distances. 

%%%%%%%%%%%%%%% FIG 2 %%%%%%%%%%%%%%%
%%%%%%%%%%%%%%% FIG 2 %%%%%%%%%%%%%%%
%
%%%%%%%%%%%%%%% FIG 3 %%%%%%%%%%%%%%%
%%%%%%%%%%%%%%% FIG 3 %%%%%%%%%%%%%%%

In the left panel of figure~\ref{fig:Pmue-color-grade}, plotted are the equi-probability contours of $P(\nu_{\mu} \rightarrow \nu_{e})$ in energy $E$ and distance $L$ plane, displayed by color-graded regions as well as the contours superimposed on it. It presents a global view of $P(\nu_{\mu} \rightarrow \nu_{e})$ which serves for the pedagogical discussions as given in section~\ref{sec:introduction}. It illuminates, in particular, the two enhanced regions of the solar- and the atmospheric-scale MSW enhancement. 
Whereas in the right panel of figure~\ref{fig:Pmue-color-grade}, presented is the ratio $\left[ P(\nu_{\mu} \rightarrow \nu_{e})^{(0)} - P(\nu_{\mu} \rightarrow \nu_{e})_{ \text{exact}  } \right] / P(\nu_{\mu} \rightarrow \nu_{e})_{ \text{exact}  }$ to show how good is the leading order formula. Or, more precisely, it is meant to be a measure for revealing where is the applicability region of our framework. One can see that the region of agreement, the left-sided yellow colored region, essentially coincides with the region of solar-scale enhancement, the target region of our perturbative treatment, as it should be. 

Though not shown, the agreement is much better in the disappearance channels, $P(\nu_{e} \rightarrow \nu_{e})$ and $P(\nu_{\mu} \rightarrow \nu_{\mu})$, $\lsim$ 1\% deviation from the exact results. For $P(\nu_{e} \rightarrow \nu_{\tau})$ the agreement is slightly better than $P(\nu_{\mu} \rightarrow \nu_{e})$. 

While the region of validity of our framework essentially traces the resonance region, there is a visible departure, or widening, of the (yellow-white) validity region toward shorter baseline to $L \sim 100$ km at low energies $E \lsim 400$ MeV. It has an interesting consequence that our formula may be applicable for shorter baseline, e.g., the T2K/T2HK baseline, $L=300$ km. Then, we should try to understand theoretically what would be the condition for validity of our formula.

Our approach toward this understanding is still ``empirical'' one, which starts from the right panel of figure~\ref{fig:Pmue-color-grade}. By approaching from the higher-energy right blue region, we encounter a line which can roughly be expressed as 
\begin{eqnarray} 
L \gsim \ell_{ \text{atm} } \equiv \frac{ 4\pi E}{ \Delta m^2_{31} } 
\label{boundary-validity}
\end{eqnarray}
for the region of validity. That is, at distance scale much larger than $\ell_{ \text{atm} }$, the neutrino evolution is insensitive to short-wavelength atmospheric oscillations, and hence our perturbative formula applies. 
On the contrary, in the alternative region $\frac{L}{E} \lsim \frac{ 4\pi }{ \Delta m^2_{31} }$, the energy scale is so high for neutrinos such that it can probe the $\Delta m^2_{31}$ effect, which renders approximation of our perturbative theory inaccurate. For this reason our leading order perturbative formula ceases to be a good approximation in region $L \lsim \ell_{ \text{atm} } $.

In figure~\ref{fig:Pmue-mumu}, comparisons are made between the energy dependences of $P(\nu_{\mu} \rightarrow \nu_{e})$ (upper cluster of four panels) and $P(\nu_{\mu} \rightarrow \nu_{\mu})$ (lower cluster of four panels) calculated with our perturbative formulas and the exact numerical computation. The leading order approximation $P(\nu_{\mu} \rightarrow \nu_{\alpha})^{(0)}$ and the one to first order correction $P(\nu_{\mu} \rightarrow \nu_{\alpha})^{(0)} + P(\nu_{\mu} \rightarrow \nu_{\alpha})^{(1)}$ ($\alpha= e, \mu$) are given by the green dotted and the red dashed lines, respectively. Whereas, the results of exact calculation are given by the blue solid lines. In each cluster in figure~\ref{fig:Pmue-mumu}, the four panels are for distances $L=300$ km, 1000 km, 2000 km, and 10000 km. 

Both in the appearance and disappearance channels, $P(\nu_{\mu} \rightarrow \nu_{e})$ and $P(\nu_{\mu} \rightarrow \nu_{\mu})$, the red dashed lines (the zeroth plus first order formulas) perfectly overlap with the blue solid lines, the exact results, to $1\%$ level or below in the region displayed in figure~\ref{fig:Pmue-mumu}. In the appearance channel the zeroth-order formula shows some deviation in region $E \gsim 300$ MeV. But, in the disappearance channel even the zeroth-order formula overlaps well with the exact result. The careful readers might have detected a tiny deviation of the first order formulas from the exact result at the peaks and at the bottoms of dips at relatively higher energies, $E \gsim 500$ MeV. 
Therefore, our perturbative framework works as expected in the target region of our formulation, $E \simeq \mbox{a few} \times 100$ MeV and $L \simeq \mbox{a few} \times 1000$ km. We also found that our first order formula $P(\nu_{\mu} \rightarrow \nu_{e})^{(0)} + P(\nu_{\mu} \rightarrow \nu_{e})^{(1)}$ works very well at low energies for the T2K/T2HK baseline, $L=300$ km.

\section{Concluding remarks}
\label{sec:conclusion}

In this paper, we have constructed a perturbative framework, the ``solar resonance perturbation theory'', whose leading order describes the solar-scale MSW enhancement in a good approximation. We aim at illuminating physics of atmospheric or accelerator neutrinos for the regions $E \simeq \mbox{a few} \times 100$ MeV and $L \simeq \mbox{a few} \times 1000$ km. Despite the keen interests expressed by people and good amount of works done \cite{Peres:2003wd,Peres:2009xe,Akhmedov:2008qt,Razzaque:2014vba}, it appears to us that experimental exploration of the physics in this region has not been performed in a sufficient depth. We hope that our discussion at a qualitative level may be useful to trigger interests of wider class of people in the community.

The clearest message which is born out from our treatment is that the effect of CP violating phase is large, larger by a factor of $\sim$10 than that expected in the conventional LBL neutrino experiments such as T2HK and DUNE, which utilize the atmospheric oscillation maximum. The CP phase effect is large at around the solar resonance or oscillation maximum because there is no suppression factor $\frac{\Delta_{21} x}{ 2 } \simeq \Delta m^2_{21} / \Delta m^2_{31} = \epsilon$ which does exist at around the atmospheric oscillation maximum. Instead, $\sin \frac{\Delta_{21} x}{ 2 } \simeq 1$ at around the solar oscillation maximum. 

Probably, the most practical way of exploring the region of solar resonance / oscillation maximum is to observe atmospheric neutrinos at low energies, ideally $E \lsim 500$ MeV. This possibility is becoming more and more feasible because of construction of massive detectors such as Hyper-K \cite{Abe:2015zbg}, DUNE \cite{Acciarri:2015uup}, JUNO \cite{An:2015jdp} (see \cite{Mari-etal-Nutele2019}), and INO \cite{Kumar:2017sdq}. It is worth to watch to what extent the energy threshold of gigantic size neutrino detectors, IceCube-Gen2/PINGU and KM3NeT-ORCA \cite{TheIceCube-Gen2:2016cap,Adrian-Martinez:2016zzs}, can be lowered. Yet, the major issues here seems to be how to improve accuracies of the atmospheric neutrino flux prediction, and reduce uncertainties in cross section measurement \cite{Okumura-private}. In this context, it is very interesting to know precisely the performance of the large-volume liquid scintillator detector JUNO for the detection of low energy atmospheric neutrinos \cite{Mari-etal-Nutele2019}. 

Our framework and the resulting oscillation formulas based on constant matter density approximation might have more direct relevance if accelerator LBL neutrino experiment is a viable possibility to explore this region. It is not only because the uniform density approximation may apply but also because it enables one to fully enjoy maximal CP violating effect by tuning the beam energy and baseline to the solar-scale enhanced region. 
Probably, a possible setup closest to the reality may be to use JPARC neutrino beam pointed to the second Hyper-K detector in Korea, the T2HKK project \cite{Abe:2016ero}, which has the baseline $L\simeq 1100$ km. We note that more than 1/3 of the $\nu_{e}$ appearance signals are in low energy bins, $E \lsim 600$ MeV. See Fig.~13 in ref.~\cite{Abe:2016ero}. Therefore, it might be worthwhile to look at this possibility in detail. 
Interestingly, our first-order perturbative formula for the oscillation probability works very well for T2K/T2HK and T2HKK setups in the relevant region of energy and baseline, as we saw in section~\ref{sec:how-accurate}.

We are aware that our methodological view for determination of the remaining unknowns in the lepton mixing and the neutrino mass pattern described in section~\ref{sec:introduction} is too biased. For example, JUNO \cite{An:2015jdp} tries to determine the mass ordering without recourse to the matter effect. Before reactor neutrino measurement of $\theta_{13}$ became popular, the majority of people thought about using accelerator neutrino beam to determine $\theta_{13}$, and essentially nobody talked about precision measurement of $\theta_{13}$ by reactor neutrinos \cite{Minakata:2002jv}. This example illuminates the danger we might have when we rely too much on the common knowledge, or a prejudice. In this sense it is worthwhile, we believe, to develop {\em alternative methods} for determining the unknowns in lepton flavor mixing. It also should have merit in the context of paradigm test (see e.g., \cite{Antusch:2006vwa,Escrihuela:2015wra,Fong:2016yyh,Fong:2017gke}), given that the effect of unitarity violation is large in the target region of this paper \cite{Fong:2017gke}. The direction we described in this paper is just one of the alternative possibilities. 

\acknowledgments

One of the authors (H.M.) thanks Kimihiro Okumura, Morihiro Honda, and Hiroshi Nunokawa for illuminating discussions on physics of low energy atmospheric neutrinos. He is grateful to Director Takaaki Kajita of Institute for Cosmic Ray Research, University of Tokyo, for hospitality in his five months stay in Research Center for Cosmic Neutrinos, where this work was initiated, expanded, and then essentially completed. 

The other author (I.M.S.) acknowledges travel support from the Colegio de F\'isica Fundamental e Interdisciplinaria de las Am\'ericas (COFI). Fermilab is operated by the Fermi Research Alliance, LLC under contract No.~DE-AC02-07CH11359 with the United States Department of Energy.

\appendix

\section{Flavor basis $S$ matrix}
\label{sec:S-matrix}

For convenience of nomenclature we define another intermediate basis, the ``tilde'' basis. The definition of $\tilde{S}$ matrix and its relation to $S$ matrix are given by 
\begin{eqnarray}
\tilde{S} \equiv 
U_{13} U_{\varphi} \hat{S} U_{\varphi}^{\dagger} U_{13}^{\dagger}, 
\hspace{10mm}
S(x) = U_{23} \tilde{S} (x) U_{23}^{\dagger}. 
\label{tildeS-S-relation}
\end{eqnarray}
Using (\ref{hat-S-matrix-1st}), the explicit expressions of $\tilde{S}$ matrix elements are given by 
\begin{eqnarray} 
\tilde{S}_{ee} &=& 
c^2_{13} \left( c^2_{\varphi} e^{ - i h_{1} x } + s^2_{\varphi} e^{ - i h_{2} x } \right) 
+ s^2_{13} e^{ - i h_{3} x }
\nonumber \\ 
&+&
2 c^2_{13} s^2_{13} 
\left\{
\frac{ \Delta_{a} }{ h_{3} - h_{1} } 
c^2_{\varphi} \left( e^{ - i h_{3} x } - e^{ - i h_{1} x } \right) 
+ \frac{ \Delta_{a} }{ h_{3} - h_{2} } 
s^2_{\varphi} \left( e^{ - i h_{3} x } - e^{ - i h_{2} x } \right) 
\right\}, 
\nonumber \\
%%%%%%%%%%%%%%%%%%%%%%%%%%%%%%%%%%%%%
\tilde{S}_{e \mu} &=& 
c_{13} c_{\varphi} s_{\varphi} \left( e^{ - i h_{2} x } - e^{ - i h_{1} x } \right) 
\nonumber \\
&-&
c_{13} s^2_{13} c_{\varphi} s_{\varphi} 
\left\{ 
\frac{ \Delta_{a} }{ h_{3} - h_{1} } 
\left( e^{ - i h_{3} x } - e^{ - i h_{1} x } \right) 
- \frac{ \Delta_{a} }{ h_{3} - h_{2} } 
\left( e^{ - i h_{3} x } - e^{ - i h_{2} x } \right) 
\right\}, 
\nonumber \\ 
%%%%%%%%%%%%%%%%%%%%%%%%%%%%%%%%%%%%%
\tilde{S}_{e \tau} &=& 
- c_{13} s_{13}  
\left\{
\left( c^2_{\varphi} e^{ - i h_{1} x } + s^2_{\varphi} e^{ - i h_{2} x } \right) - e^{ - i h_{3} x } 
\right\}
\nonumber \\ 
&+&
( c^2_{13} - s^2_{13} ) c_{13} s_{13} 
\left\{
c^2_{\varphi} \frac{ \Delta_{a} }{ h_{3} - h_{1} } 
\left( e^{ - i h_{3} x } - e^{ - i h_{1} x } \right) 
+ s^2_{\varphi} \frac{ \Delta_{a} }{ h_{3} - h_{2} } 
\left( e^{ - i h_{3} x } - e^{ - i h_{2} x } \right)
\right\}, 
\nonumber \\
%%%%%%%%%%%%%%%%%%%%%%%%%%%%%%%%%%%
\tilde{S}_{\mu e} &=& 
c_{13} c_{\varphi} s_{\varphi} \left( e^{ - i h_{2} x } - e^{ - i h_{1} x } \right) 
\nonumber \\
&-&
c_{13} s^2_{13} c_{\varphi} s_{\varphi} 
\left\{ \frac{ \Delta_{a} }{ h_{3} - h_{1} } 
\left( e^{ - i h_{3} x } - e^{ - i h_{1} x } \right) 
- \frac{ \Delta_{a} }{ h_{3} - h_{2} } 
\left( e^{ - i h_{3} x } - e^{ - i h_{2} x } \right) \right\} 
= \tilde{S}_{e \mu}, 
\nonumber \\
%%%%%%%%%%%%%%%%%%%%%%%%%%%%%%%%%%%
\tilde{S}_{\mu \mu} &=& 
\left( s^2_{\varphi} e^{ - i h_{1} x } + c^2_{\varphi} e^{ - i h_{2} x } \right),
\nonumber \\
%%%%%%%%%%%%%%%%%%%%%%%%%%%%%%%%%%%
\tilde{S}_{\mu \tau} &=& 
- s_{13} c_{\varphi} s_{\varphi} \left( e^{ - i h_{2} x } - e^{ - i h_{1} x } \right) 
\nonumber \\
&-&
c^2_{13} s_{13} c_{\varphi} s_{\varphi} 
\left\{ \frac{ \Delta_{a} }{ h_{3} - h_{1} } 
\left( e^{ - i h_{3} x } - e^{ - i h_{1} x } \right) 
- \frac{ \Delta_{a} }{ h_{3} - h_{2} } 
\left( e^{ - i h_{3} x } - e^{ - i h_{2} x } \right) \right\},
\nonumber \\ 
%%%%%%%%%%%%%%%%%%%%%%%%%%%%%%%%%%%
\tilde{S}_{\tau e} &=& 
- c_{13} s_{13} 
\left\{
\left( c^2_{\varphi} e^{ - i h_{1} x } + s^2_{\varphi} e^{ - i h_{2} x } \right) - e^{ - i h_{3} x } 
\right\} 
\nonumber \\
&+&
( c^2_{13} - s^2_{13} ) c_{13} s_{13} 
\left\{
c^2_{\varphi} \frac{ \Delta_{a} }{ h_{3} - h_{1} } 
\left( e^{ - i h_{3} x } - e^{ - i h_{1} x } \right) 
+ s^2_{\varphi} \frac{ \Delta_{a} }{ h_{3} - h_{2} } 
\left( e^{ - i h_{3} x } - e^{ - i h_{2} x } \right)
\right\} 
= \tilde{S}_{e \tau},
\nonumber \\ 
%%%%%%%%%%%%%%%%%%%%%%%%%%%%%%%%%%%
\tilde{S}_{\tau \mu} &=& 
- s_{13} c_{\varphi} s_{\varphi} \left( e^{ - i h_{2} x } - e^{ - i h_{1} x } \right) 
\nonumber \\
&-&
c^2_{13} s_{13} c_{\varphi} s_{\varphi} 
\left\{ \frac{ \Delta_{a} }{ h_{3} - h_{1} } 
\left( e^{ - i h_{3} x } - e^{ - i h_{1} x } \right) 
- \frac{ \Delta_{a} }{ h_{3} - h_{2} } 
\left( e^{ - i h_{3} x } - e^{ - i h_{2} x } \right) \right\} 
= \tilde{S}_{\mu \tau}, 
\nonumber \\
%%%%%%%%%%%%%%%%%%%%%%%%%%%%%%%%%%%
\tilde{S}_{\tau \tau} &=& 
s^2_{13} 
\left( c^2_{\varphi} e^{ - i h_{1} x } + s^2_{\varphi} e^{ - i h_{2} x } \right) 
+ c^2_{13} 
e^{ - i h_{3} x } 
\nonumber \\
&-&
2 c^2_{13} s^2_{13} 
\left\{
c^2_{\varphi} \frac{ \Delta_{a} }{ h_{3} - h_{1} } 
\left( e^{ - i h_{3} x } - e^{ - i h_{1} x } \right) 
+ s^2_{\varphi} \frac{ \Delta_{a} }{ h_{3} - h_{2} } 
\left( e^{ - i h_{3} x } - e^{ - i h_{2} x } \right) 
\right\}.
\label{tildeS-elements-final}
\end{eqnarray}
Then, finally, $U_{23}$ rotation can be performed to convert the above expression of $\tilde{S}_{\alpha \beta}$ to $S$ matrix elements: 
\begin{eqnarray} 
S_{ee} &=& \tilde{ S }_{ee}, 
\nonumber \\
S_{e \mu} &=& c_{23} \tilde{ S }_{e \mu} + s_{23} e^{ - i \delta} \tilde{ S }_{e \tau}, 
\nonumber \\ 
S_{e \tau} &=& c_{23} \tilde{ S }_{e \tau} - s_{23} e^{ i \delta} \tilde{ S }_{e \mu},
\nonumber \\
S_{\mu e} &=& c_{23} \tilde{ S }_{\mu e} + s_{23} e^{ i \delta} \tilde{ S }_{\tau e} 
= S_{e \mu} (- \delta),  
\nonumber \\
S_{\mu \mu} &=& c^2_{23} \tilde{ S }_{\mu \mu} + s^2_{23} \tilde{ S }_{\tau \tau} + c_{23} s_{23} ( e^{ - i \delta} \tilde{ S }_{\mu \tau} + e^{ i \delta}  \tilde{ S }_{\tau \mu} ), 
\nonumber \\
S_{\mu \tau} &=& c^2_{23} \tilde{ S }_{\mu \tau} - s^2_{23} e^{ 2 i \delta}  \tilde{ S }_{\tau \mu} + c_{23} s_{23} e^{ i \delta}  ( \tilde{ S }_{\tau \tau} - \tilde{ S }_{\mu \mu} ), 
\nonumber \\ 
S_{\tau e} &=& c_{23} \tilde{ S }_{\tau e} - s_{23} e^{ - i \delta} \tilde{ S }_{\mu e} 
= S_{e \tau} (- \delta), 
\nonumber \\ 
S_{\tau \mu} &=& c^2_{23} \tilde{ S }_{\tau \mu } - s^2_{23} e^{ - 2 i \delta} \tilde{ S }_{ \mu \tau} + c_{23} s_{23} e^{ - i \delta} ( \tilde{ S }_{\tau \tau} - \tilde{ S }_{\mu \mu} ) = S_{\mu \tau} (- \delta),
\nonumber \\
S_{\tau \tau} &=& s^2_{23} \tilde{ S }_{\mu \mu} + c^2_{23} \tilde{ S }_{\tau \tau} - c_{23} s_{23} ( e^{ - i \delta} \tilde{ S }_{\mu \tau} + e^{ i \delta} \tilde{ S }_{\tau \mu} ).
\label{S-matrix-elements}
\end{eqnarray}

\section{The expressions of neutrino oscillation probability}
\label{sec:oscillation-probability}

The oscillation probability can be calculated as 
\begin{eqnarray} 
P(\nu_{\beta} \rightarrow \nu_{\alpha}; x) = \vert S_{\alpha \beta} \vert^2.
\label{oscillation-probability-def}
\end{eqnarray}
We calculate $P(\nu_{\beta} \rightarrow \nu_{\alpha}; x)$ to first order in the effective expansion parameter, and the results are presented in the following way:
\begin{eqnarray} 
P(\nu_{\beta} \rightarrow \nu_{\alpha}) = 
P(\nu_{\beta} \rightarrow \nu_{\alpha})^{(0)}
+ P(\nu_{\beta} \rightarrow \nu_{\alpha})^{(1)}.
\label{oscillation-probability-general}
\end{eqnarray}
The explicit expressions of the leading-order term, $P(\nu_{\beta} \rightarrow \nu_{\alpha})^{(0)}$, and the first-order corrections $P(\nu_{\beta} \rightarrow \nu_{\alpha})^{(1)}$ are given below. 

\subsection{The oscillation probability in the $\nu_{e}$ sector: zeroth-order}
\label{sec:probability-nue-0th}

The reduced Jarlskog factor in matter is defined in eq.~(\ref{Jmr-def}) as $J_{mr} \equiv c_{23} s_{23} c^2_{13} s_{13} c_{\varphi} s_{\varphi}$. The oscillation probabilities in $\nu_{e}$ sector, 
$P(\nu_{e} \rightarrow \nu_{e})^{(0)}$, 
$P(\nu_{e} \rightarrow \nu_{\mu})^{(0)}$, and 
$P(\nu_{e} \rightarrow \nu_{\tau})^{(0)}$ are given as: 
\begin{eqnarray} 
&& 
P(\nu_{e} \rightarrow \nu_{e})^{(0)} 
\nonumber \\
&=& 
1 - c^4_{13} \sin^2 2 \varphi \sin^2 \frac{ ( h_{2} - h_{1} ) x }{2}
- \sin^2 2\theta_{13}  
\left[
c^2_{\varphi} \sin^2 \frac{ ( h_{3} - h_{1} ) x }{2} 
+ s^2_{\varphi} \sin^2 \frac{ ( h_{3} - h_{2} ) x }{2} 
\right].
\nonumber \\
\label{P-ee-0th}
\end{eqnarray}
\begin{eqnarray} 
&& 
P(\nu_{e} \rightarrow \nu_{\mu})^{(0)} =
c^2_{23} c^2_{13} \sin^2 2 \varphi 
\sin^2 \frac{ ( h_{2} - h_{1} ) x }{2}
\nonumber \\
&+& 
s^2_{23} \sin^2 2\theta_{13} 
\left[
c^2_{\varphi} \sin^2 \frac{ ( h_{3} - h_{1} ) x }{2} 
+ s^2_{\varphi} \sin^2 \frac{ ( h_{3} - h_{2} ) x }{2} 
- c^2_{\varphi} s^2_{\varphi} \sin^2 \frac{ ( h_{2} - h_{1} ) x }{2} 
\right]
\nonumber \\
&+& 
%%c_{23} s_{23} c^2_{13} s_{13} c_{\varphi} s_{\varphi} 
4 J_{mr} \cos \delta 
\left\{
\cos 2 \varphi \sin^2 \frac{ ( h_{2} - h_{1} ) x }{2} 
- \sin^2 \frac{ ( h_{3} - h_{2} ) x }{2} 
+ \sin^2 \frac{ ( h_{3} - h_{1} ) x }{2} 
\right\} 
\nonumber \\
&-& 
8 J_{mr} \sin \delta 
\sin \frac{ ( h_{3} - h_{2} ) x }{2} 
\sin \frac{ ( h_{2} - h_{1} ) x }{2} 
\sin \frac{ ( h_{1} - h_{3} ) x }{2}. 
\label{P-emu-0th}
\end{eqnarray}
\begin{eqnarray} 
&& 
P(\nu_{e} \rightarrow \nu_{\tau})^{(0)}
\nonumber \\
&=& 
c^2_{23} \sin^2 2\theta_{13} %% 4 c^2_{13} s^2_{13} 
\left[
- c^2_{\varphi} s^2_{\varphi} 
%\frac{1}{4} \sin^2 2 \varphi 
\sin^2 \frac{ ( h_{2} - h_{1} ) x }{2} 
+ c^2_{\varphi} \sin^2 \frac{ ( h_{3} - h_{1} ) x }{2} 
+ s^2_{\varphi} \sin^2 \frac{ ( h_{3} - h_{2} ) x }{2}
\right]
\nonumber \\
&+&
s^2_{23} c^2_{13} \sin^2 2 \varphi \sin^2 \frac{ ( h_{2} - h_{1} ) x }{2} 
\nonumber \\
&-&
4 J_{mr} \cos \delta 
\left\{
\cos 2 \varphi \sin^2 \frac{ ( h_{2} - h_{1} ) x }{2} 
- \sin^2 \frac{ ( h_{3} - h_{2} ) x }{2} 
+ \sin^2 \frac{ ( h_{3} - h_{1} ) x }{2} 
\right\} 
\nonumber \\
&+& 
8 J_{mr} \sin \delta 
\sin \frac{ ( h_{3} - h_{2} ) x }{2} 
\sin \frac{ ( h_{2} - h_{1} ) x }{2} 
\sin \frac{ ( h_{1} - h_{3} ) x }{2}. 
\label{P-etau-0th}
\end{eqnarray}
The unitarity in $\nu_{e}$ row at zeroth order,
\begin{eqnarray} 
P(\nu_{e} \rightarrow \nu_{e})^{(0)} 
+ P(\nu_{e} \rightarrow \nu_{\mu})^{(0)}
+ P(\nu_{e} \rightarrow \nu_{\tau})^{(0)} = 1,
\end{eqnarray}
can be checked by inspection.

\subsection{The oscillation probability in the $\nu_{\mu} - \nu_{\tau}$ sector: zeroth-order}
\label{sec:probability-numu-nutau-0th}

$P(\nu_{\mu} \rightarrow \nu_{e})^{(0)}$ and $P(\nu_{\mu} \rightarrow \nu_{\mu})^{(0)}$ are given in eqs.~(\ref{P-mue-0th}), and (\ref{P-mumu-0th}), respectively, and hence we do not repeat. $P(\nu_{\mu} \rightarrow \nu_{\tau})^{(0)}$ are given by 
\begin{eqnarray} 
&& 
P(\nu_{\mu} \rightarrow \nu_{\tau})^{(0)} 
\nonumber \\
&=& 
4 \left[ 
s^2_{13} \left( c^2_{\varphi} s^2_{\varphi} + c^2_{23} s^2_{23} \right) 
- c^2_{23} s^2_{23} c^2_{\varphi} s^2_{\varphi} \left( 1 + 4 s^2_{13} + s^4_{13} \right)
\right]
\sin^2 \frac{ ( h_{2} - h_{1} ) x }{2} 
\nonumber \\
&-& 
4 c^2_{23} s^2_{23} c^2_{13} 
\left[ 
( s^2_{13} c^2_{\varphi} - s^2_{\varphi} )
\sin^2 \frac{ ( h_{3} - h_{1} ) x }{2} 
+ 
( s^2_{13} s^2_{\varphi} - c^2_{\varphi} ) 
\sin^2 \frac{ ( h_{3} - h_{2} ) x }{2} 
\right]
\nonumber \\
&-&
8 c^2_{23} s^2_{23} s^2_{13} c^2_{\varphi} s^2_{\varphi} \cos 2\delta 
\sin^2 \frac{ ( h_{2} - h_{1} ) x }{2} 
\nonumber \\
&+& 
4 c_{23} s_{23} s_{13} c_{\varphi} s_{\varphi} \cos 2\theta_{23} \cos \delta 
\nonumber \\
&\times&
\left[
\left( 1 + s^2_{13} \right) \cos 2\varphi 
\sin^2 \frac{ ( h_{2} - h_{1} ) x }{2} 
+
c^2_{13} 
\left\{
\sin^2 \frac{ ( h_{3} - h_{2} ) x }{2} - \sin^2 \frac{ ( h_{3} - h_{1} ) x }{2} 
\right\}
\right]
\nonumber \\
&-& 
8 J_{mr}
%%c_{23} s_{23} c^2_{13} s_{13} c_{\varphi} s_{\varphi} 
\sin \delta 
\sin \frac{ ( h_{3} - h_{2} ) x }{2} 
\sin \frac{ ( h_{2} - h_{1} ) x }{2} 
\sin \frac{ ( h_{1} - h_{3} ) x }{2}. 
\label{P-mutau-0th}
\end{eqnarray}
Using these expressions, one can easily verify unitarity in $\nu_{\mu}$ row at zeroth order,  
\begin{eqnarray} 
P(\nu_{\mu} \rightarrow \nu_{e})^{(0)} 
+ P(\nu_{\mu} \rightarrow \nu_{\mu})^{(0)}
+ P(\nu_{\mu} \rightarrow \nu_{\tau})^{(0)} = 1.
\end{eqnarray}

\subsection{The oscillation probability in the $\nu_{e}$ sector: first order corrections}
\label{sec:probability-nue-1st}

The first-order corrections to the oscillation probability in the $\nu_{e}$ sector are given by:
\begin{eqnarray} 
&& 
P(\nu_{e} \rightarrow \nu_{e})^{(1)} 
\nonumber \\
&=&
2 \sin^2 2\theta_{13} 
\frac{ \Delta_{a} }{ h_{3} - h_{1} } 
\biggl[ 
c^2_{\varphi} \left( s^2_{13} - c^2_{13} c^2_{\varphi} \right) 
\sin^2 \frac{ ( h_{3} - h_{1} ) x }{2}
+ 
c^2_{13} c^2_{\varphi} s^2_{\varphi} 
\left\{
\sin^2 \frac{ ( h_{2} - h_{1} ) x }{2} - \sin^2 \frac{ ( h_{3} - h_{2} ) x }{2} 
\right\}
\biggr]
\nonumber \\
&+&
2 \sin^2 2\theta_{13} 
\frac{ \Delta_{a} }{ h_{3} - h_{2} } 
\biggl[ 
s^2_{\varphi} \left( s^2_{13} - c^2_{13} s^2_{\varphi} \right) 
\sin^2 \frac{ ( h_{3} - h_{2} ) x }{2}
+ 
c^2_{13} c^2_{\varphi} s^2_{\varphi} 
\left\{
\sin^2 \frac{ ( h_{2} - h_{1} ) x }{2} - \sin^2 \frac{ ( h_{3} - h_{1} ) x }{2} 
\right\}
\biggr].
\nonumber \\
\label{P-ee-1st}
\end{eqnarray}
\begin{eqnarray} 
&& 
P(\nu_{e} \rightarrow \nu_{\mu})^{(1)} 
\nonumber \\ 
&=& 
4 c^2_{13} s^2_{13} c^2_{\varphi} \frac{ \Delta_{a} }{ h_{3} - h_{1} } 
\biggl[ 
\left\{ s^2_{23} \cos 2\theta_{13} ( 1 + c^2_{\varphi} ) - c^2_{23} s^2_{\varphi} 
\right\}
\sin^2 \frac{ ( h_{3} - h_{1} ) x }{2} 
\nonumber \\
&-& 
s^2_{\varphi} 
\left( c^2_{23} + s^2_{23} \cos 2\theta_{13} \right) 
\left\{ \sin^2 \frac{ ( h_{2} - h_{1} ) x }{2} - \sin^2 \frac{ ( h_{3} - h_{2} ) x }{2} \right\}
\biggr]
\nonumber \\
&+& 
4 c^2_{13} s^2_{13} s^2_{\varphi} \frac{ \Delta_{a} }{ h_{3} - h_{2} }  
\biggl[ 
\left\{ 
s^2_{23} \cos 2\theta_{13} ( 1 + s^2_{\varphi} ) - c^2_{23} c^2_{\varphi} 
\right\}
\sin^2 \frac{ ( h_{3} - h_{2} ) x }{2} 
\nonumber \\
&-& 
c^2_{\varphi} \left( c^2_{23} + s^2_{23} \cos 2\theta_{13} \right) 
\left\{ \sin^2 \frac{ ( h_{2} - h_{1} ) x }{2} - \sin^2 \frac{ ( h_{3} - h_{1} ) x }{2} \right\}
\biggr] 
\nonumber \\
&+& 
4 J_{mr} \cos \delta
%% c_{23} s_{23} c^2_{13} s_{13} c_{\varphi} s_{\varphi} 
\frac{ \Delta_{a} }{ h_{3} - h_{1} } 
\biggl[
\left\{ \cos 2\theta_{13} c^2_{\varphi} - s^2_{13} ( 1 + c^2_{\varphi} ) 
\right\} 
\sin^2 \frac{ ( h_{3} - h_{1} ) x }{2} 
\nonumber \\
&+& 
\left( \cos 2\theta_{13} c^2_{\varphi} + s^2_{13} s^2_{\varphi}  \right) 
\left\{ \sin^2 \frac{ ( h_{2} - h_{1} ) x }{2} - \sin^2 \frac{ ( h_{3} - h_{2} ) x }{2} \right\} 
\biggr]
\nonumber \\
&-& 
4 J_{mr} \cos \delta
%%c_{23} s_{23} c^2_{13} s_{13} c_{\varphi} s_{\varphi} 
\frac{ \Delta_{a} }{ h_{3} - h_{2} } 
\biggl[
\left\{ \cos 2\theta_{13} s^2_{\varphi} - s^2_{13} ( 1 + s^2_{\varphi} ) 
\right\} 
\sin^2 \frac{ ( h_{3} - h_{2} ) x }{2} 
\nonumber \\
&+& 
\left( \cos 2\theta_{13} s^2_{\varphi} + s^2_{13} c^2_{\varphi}  \right) 
\left\{ \sin^2 \frac{ ( h_{2} - h_{1} ) x }{2} - \sin^2 \frac{ ( h_{3} - h_{1} ) x }{2} \right\} 
\biggr]
\nonumber \\
&+& 
8 J_{mr} \sin \delta
%%c_{23} s_{23} c^2_{13} s_{13} c_{\varphi} s_{\varphi} 
\left( s^2_{13} - c^2_{13} c^2_{\varphi} \right) 
\frac{ \Delta_{a} }{ h_{3} - h_{1} } 
\sin \frac{ ( h_{3} - h_{2} ) x }{2} 
\sin \frac{ ( h_{2} - h_{1} ) x }{2} 
\sin \frac{ ( h_{1} - h_{3} ) x }{2} 
\nonumber \\
&+& 
8 J_{mr} \sin \delta
%%c_{23} s_{23} c^2_{13} s_{13} c_{\varphi} s_{\varphi} 
\left( s^2_{13} - c^2_{13} s^2_{\varphi}  \right) 
\frac{ \Delta_{a} }{ h_{3} - h_{2} } 
\sin \frac{ ( h_{3} - h_{2} ) x }{2} 
\sin \frac{ ( h_{2} - h_{1} ) x }{2} 
\sin \frac{ ( h_{1} - h_{3} ) x }{2}.
\nonumber \\
\label{P-emu-1st}
\end{eqnarray}
\begin{eqnarray} 
&& 
P(\nu_{e} \rightarrow \nu_{\tau})^{(1)} 
\nonumber \\ 
&=& 
4 c^2_{13} s^2_{13} c^2_{\varphi} 
\frac{ \Delta_{a} }{ h_{3} - h_{1} } 
\biggl[ 
\left\{ c^2_{23} \cos 2\theta_{13} ( 1 + c^2_{\varphi} ) - s^2_{23} s^2_{\varphi} 
\right\}
\sin^2 \frac{ ( h_{3} - h_{1} ) x }{2} 
\nonumber \\
&+& 
s^2_{\varphi} 
\left( s^2_{23} + c^2_{23} \cos 2\theta_{13} \right) 
\left\{ \sin^2 \frac{ ( h_{3} - h_{2} ) x }{2} - \sin^2 \frac{ ( h_{2} - h_{1} ) x }{2} \right\}
\biggr]
\nonumber \\
&+& 
4 c^2_{13} s^2_{13} s^2_{\varphi} 
\frac{ \Delta_{a} }{ h_{3} - h_{2} } 
\biggl[ 
\left\{ c^2_{23} \cos 2\theta_{13} ( 1 + s^2_{\varphi} ) - s^2_{23} c^2_{\varphi} 
\right\}
\sin^2 \frac{ ( h_{3} - h_{2} ) x }{2} 
\nonumber \\
&+& 
c^2_{\varphi} 
\left( s^2_{23} + c^2_{23} \cos 2\theta_{13} \right) 
\left\{ \sin^2 \frac{ ( h_{3} - h_{1} ) x }{2} - \sin^2 \frac{ ( h_{2} - h_{1} ) x }{2} \right\}
\biggr]
\nonumber \\
&+& 
4 J_{mr} \cos \delta 
%%c_{23} s_{23} c^2_{13} s_{13} c_{\varphi} s_{\varphi} 
\frac{ \Delta_{a} }{ h_{3} - h_{1} } 
\biggl[
\left\{ s^2_{13} ( 1 + c^2_{\varphi} ) - \cos 2\theta_{13} c^2_{\varphi} \right\} 
\sin^2 \frac{ ( h_{3} - h_{1} ) x }{2} 
\nonumber \\
&+& 
\left( s^2_{13} s^2_{\varphi} + \cos 2\theta_{13} c^2_{\varphi} \right) 
\left\{ \sin^2 \frac{ ( h_{3} - h_{2} ) x }{2} - \sin^2 \frac{ ( h_{2} - h_{1} ) x }{2} \right\} 
\biggr]
\nonumber \\
&-& 
4 J_{mr} \cos \delta 
%%c_{23} s_{23} c^2_{13} s_{13} c_{\varphi} s_{\varphi} 
\frac{ \Delta_{a} }{ h_{3} - h_{2} } 
\biggl[
\left\{ s^2_{13} ( 1 + s^2_{\varphi} ) - \cos 2\theta_{13} s^2_{\varphi} \right\} 
\sin^2 \frac{ ( h_{3} - h_{2} ) x }{2} 
\nonumber \\
&+& 
\left( s^2_{13} c^2_{\varphi} + \cos 2\theta_{13} s^2_{\varphi} \right) 
\left\{ \sin^2 \frac{ ( h_{3} - h_{1} ) x }{2} - \sin^2 \frac{ ( h_{2} - h_{1} ) x }{2} \right\} 
\biggr]
\nonumber \\
&-& 
8 J_{mr} \sin \delta
%%c_{23} s_{23} c^2_{13} s_{13} c_{\varphi} s_{\varphi} 
\left( s^2_{13} - c^2_{13} c^2_{\varphi}  \right) 
\frac{ \Delta_{a} }{ h_{3} - h_{1} } 
\sin \frac{ ( h_{3} - h_{2} ) x }{2} 
\sin \frac{ ( h_{2} - h_{1} ) x }{2} 
\sin \frac{ ( h_{1} - h_{3} ) x }{2} 
\nonumber \\
&-& 
8 J_{mr} \sin \delta
%%c_{23} s_{23} c^2_{13} s_{13} c_{\varphi} s_{\varphi} 
\left( s^2_{13} - c^2_{13} s^2_{\varphi} \right) 
\frac{ \Delta_{a} }{ h_{3} - h_{2} } 
\sin \frac{ ( h_{3} - h_{2} ) x }{2} 
\sin \frac{ ( h_{2} - h_{1} ) x }{2} 
\sin \frac{ ( h_{1} - h_{3} ) x }{2}. 
\nonumber \\
\label{P-etau-1st}
\end{eqnarray}
The unitarity in the $\nu_{e}$ row at first order, 
\begin{eqnarray} 
P(\nu_{e} \rightarrow \nu_{e})^{(1)} 
+ P(\nu_{e} \rightarrow \nu_{\mu})^{(1)}
+ P(\nu_{e} \rightarrow \nu_{\tau})^{(1)} =0, 
\end{eqnarray}
can be verified with use of eqs.~(\ref{P-ee-1st}), (\ref{P-emu-1st}), and (\ref{P-etau-1st}).

\subsection{The oscillation probability in the $\nu_{\mu} - \nu_{\tau}$ sector: first order corrections}
\label{sec:probability-mu-tau-1st}

$P(\nu_{\mu} \rightarrow \nu_{e})^{(1)}$ is given by T conjugate of $P(\nu_{e} \rightarrow \nu_{\mu})^{(1)}$ given in (\ref{P-emu-1st}). The remaining two probabilities read:
\begin{eqnarray} 
&& 
P(\nu_{\mu} \rightarrow \nu_{\mu})^{(1)} 
\nonumber \\ 
&=& 
- 8 \biggl[
s^2_{23} c^2_{13} s^2_{13} c^2_{\varphi} 
\left( s^2_{23} s^2_{13} s^2_{\varphi} + c^2_{23} \cos 2\varphi \right) 
\biggr]
%
%\nonumber \\ &\times& 
\frac{ \Delta_{a} }{ h_{3} - h_{1} } 
\biggl[
\sin^2 \frac{ ( h_{2} - h_{1} ) x }{2} 
- \sin^2 \frac{ ( h_{3} - h_{2} ) x }{2} 
\biggr]
\nonumber \\ 
&-& 
8 \biggl[
s^2_{23} c^2_{13} s^2_{13} s^2_{\varphi} 
\left( s^2_{23} s^2_{13} c^2_{\varphi} - c^2_{23} \cos 2\varphi \right) 
\biggr]
%
%\nonumber \\ &\times& 
\frac{ \Delta_{a} }{ h_{3} - h_{2} } 
\biggl[
\sin^2 \frac{ ( h_{2} - h_{1} ) x }{2} 
- \sin^2 \frac{ ( h_{3} - h_{1} ) x }{2} 
\biggr]
\nonumber \\ 
&+& 
8 \biggl[
2 c^2_{23} s^2_{23} c^2_{13} s^2_{13} c^2_{\varphi} s^2_{\varphi} 
+ s^4_{23} c^2_{13} s^2_{13} c^2_{\varphi} 
\left\{ -1 + s^2_{13} \left( 1 + c^2_{\varphi} \right) \right\} 
\biggr]
%
%\nonumber \\ &\times& 
\frac{ \Delta_{a} }{ h_{3} - h_{1} } 
\sin^2 \frac{ ( h_{3} - h_{1} ) x }{2} 
\nonumber \\ 
&+& 
8 \biggl[
2 c^2_{23} s^2_{23} c^2_{13} s^2_{13} c^2_{\varphi} s^2_{\varphi} 
+ s^4_{23} c^2_{13} s^2_{13} s^2_{\varphi} 
\left\{ -1 + s^2_{13} \left( 1 + s^2_{\varphi} \right) \right\} 
\biggr]
%
%\nonumber \\ &\times& 
\frac{ \Delta_{a} }{ h_{3} - h_{2} } 
\sin^2 \frac{ ( h_{3} - h_{2} ) x }{2} 
\nonumber \\ 
&-& 
8 
\biggl\{
\left( c^2_{23} - s^2_{23} s^2_{13} \right) c^2_{\varphi} - s^2_{23} s^2_{13} \cos 2\varphi \biggr\} J_{mr} \cos \delta 
%
%\nonumber \\ &-& c^2_{23} s^2_{23} c^2_{13} s^2_{13} c^2_{\varphi} s^2_{\varphi} \cos 2\delta 
%
%\nonumber \\ &\times& 
\frac{ \Delta_{a} }{ h_{3} - h_{1} } 
\biggl[
\sin^2 \frac{ ( h_{2} - h_{1} ) x }{2} 
- \sin^2 \frac{ ( h_{3} - h_{2} ) x }{2} 
\biggr]
\nonumber \\ 
&+& 
8 
\biggl\{
\left( c^2_{23} - s^2_{23} s^2_{13} \right) s^2_{\varphi} + s^2_{23} s^2_{13} \cos 2\varphi \biggr\} J_{mr} \cos \delta 
%
%\nonumber \\ &\times& 
\frac{ \Delta_{a} }{ h_{3} - h_{2} } 
\biggl[
\sin^2 \frac{ ( h_{2} - h_{1} ) x }{2} 
- \sin^2 \frac{ ( h_{3} - h_{1} ) x }{2} 
\biggr]
\nonumber \\ 
&+& 
8 
\biggl\{ -1 + c^2_{23} \left( 1 + s^2_{\varphi} \right) 
+ s^2_{23} s^2_{13} \left( 1 + 3 c^2_{\varphi} \right) \biggr\} J_{mr} \cos \delta 
\frac{ \Delta_{a} }{ h_{3} - h_{1} } 
\sin^2 \frac{ ( h_{3} - h_{1} ) x }{2} 
\nonumber \\ 
&-& 
8 
\biggl\{ -1 + c^2_{23} \left( 1 + c^2_{\varphi} \right) 
+ s^2_{23} s^2_{13} \left( 1 + 3 s^2_{\varphi} \right) \biggr\} J_{mr} \cos \delta 
\frac{ \Delta_{a} }{ h_{3} - h_{2} } 
\sin^2 \frac{ ( h_{3} - h_{2} ) x }{2} 
\nonumber \\ 
&+& 
8 c^2_{23} s^2_{23} c^2_{13} s^2_{13} c^2_{\varphi} s^2_{\varphi} \cos 2\delta 
\frac{ \Delta_{a} }{ h_{3} - h_{1} } 
\biggl[
\sin^2 \frac{ ( h_{2} - h_{1} ) x }{2} 
- \sin^2 \frac{ ( h_{3} - h_{2} ) x }{2} 
+ \sin^2 \frac{ ( h_{3} - h_{1} ) x }{2}
\biggr]
\nonumber \\ 
&+& 
8 c^2_{23} s^2_{23} c^2_{13} s^2_{13} c^2_{\varphi} s^2_{\varphi} \cos 2\delta 
\frac{ \Delta_{a} }{ h_{3} - h_{2} } 
\biggl[
\sin^2 \frac{ ( h_{2} - h_{1} ) x }{2} 
- \sin^2 \frac{ ( h_{3} - h_{1} ) x }{2} 
+ \sin^2 \frac{ ( h_{3} - h_{2} ) x }{2}
\biggr].
\nonumber \\
\label{P-mumu-1st}
\end{eqnarray}
\begin{eqnarray} 
&& 
P(\nu_{\mu} \rightarrow \nu_{\tau})^{(1)} 
\nonumber \\ 
&=& 
4 c^2_{13} s^2_{13} c^2_{\varphi} s^2_{\varphi} \left( 1 - 2 c^2_{23} s^2_{23} \right) 
%
%\nonumber \\ &\times& 
\frac{ \Delta_{a} }{ h_{3} - h_{1} } 
\biggl[
\sin^2 \frac{ ( h_{2} - h_{1} ) x }{2} 
- \sin^2 \frac{ ( h_{3} - h_{2} ) x }{2} 
+ \sin^2 \frac{ ( h_{3} - h_{1} ) x }{2}
\biggr]
\nonumber \\ 
&+& 
4 c^2_{13} s^2_{13} c^2_{\varphi} s^2_{\varphi} \left( 1 - 2 c^2_{23} s^2_{23} \right) 
%
%\nonumber \\ &\times& 
\frac{ \Delta_{a} }{ h_{3} - h_{2} } 
\biggl[
\sin^2 \frac{ ( h_{2} - h_{1} ) x }{2} 
+ \sin^2 \frac{ ( h_{3} - h_{2} ) x }{2} 
- \sin^2 \frac{ ( h_{3} - h_{1} ) x }{2}
\biggr]
\nonumber \\ 
&-& 
2 \sin^2 2\theta_{23} c^2_{13} s^2_{13} c^2_{\varphi} 
\frac{ \Delta_{a} }{ h_{3} - h_{1} } 
\nonumber \\ 
&\times& 
\biggl[
\left( s^2_{13} s^2_{\varphi} - c^2_{\varphi} \right) 
\left\{ \sin^2 \frac{ ( h_{2} - h_{1} ) x }{2} - \sin^2 \frac{ ( h_{3} - h_{2} ) x }{2} \right\}
+ \left\{ c^2_{13} ( 1 + c^2_{\varphi} ) - \cos 2\varphi \right\} 
\sin^2 \frac{ ( h_{3} - h_{1} ) x }{2}
\biggr]
\nonumber \\ 
&-& 
2 \sin^2 2\theta_{23} c^2_{13} s^2_{13} s^2_{\varphi} 
\frac{ \Delta_{a} }{ h_{3} - h_{2} } 
\nonumber \\ 
&\times& 
\biggl[
\left( s^2_{13} c^2_{\varphi} - s^2_{\varphi} \right) 
\left\{ \sin^2 \frac{ ( h_{2} - h_{1} ) x }{2} - \sin^2 \frac{ ( h_{3} - h_{1} ) x }{2} \right\}
+ \left\{ c^2_{13} ( 1 + s^2_{\varphi} ) + \cos 2\varphi \right\} 
\sin^2 \frac{ ( h_{3} - h_{2} ) x }{2}
\biggr]
\nonumber \\ 
&+& 
8 \left( \cos 2\theta_{23} s^2_{13} c^2_{\varphi} \right) J_{mr} \cos \delta 
%
%\nonumber \\ &\times& 
\frac{ \Delta_{a} }{ h_{3} - h_{1} } 
\biggl[
\sin^2 \frac{ ( h_{2} - h_{1} ) x }{2} 
- \sin^2 \frac{ ( h_{3} - h_{2} ) x }{2} 
+ \sin^2 \frac{ ( h_{3} - h_{1} ) x }{2}
\biggr]
\nonumber \\ 
&-& 
8 \left( \cos 2\theta_{23} s^2_{13} s^2_{\varphi} \right) J_{mr} \cos \delta
%
%\nonumber \\ &\times& 
\frac{ \Delta_{a} }{ h_{3} - h_{2} } 
\biggl[
\sin^2 \frac{ ( h_{2} - h_{1} ) x }{2} 
+ \sin^2 \frac{ ( h_{3} - h_{2} ) x }{2} 
- \sin^2 \frac{ ( h_{3} - h_{1} ) x }{2}
\biggr]
\nonumber \\ 
&-& 
4 \cos 2\theta_{23} J_{mr} \cos \delta  
\frac{ \Delta_{a} }{ h_{3} - h_{1} } 
\nonumber \\ 
&\times& 
\biggl[
\left( s^2_{13} s^2_{\varphi} - c^2_{\varphi} \right) 
\left\{ \sin^2 \frac{ ( h_{2} - h_{1} ) x }{2} - \sin^2 \frac{ ( h_{3} - h_{2} ) x }{2} \right\}
+ \left\{ c^2_{13} ( 1 + c^2_{\varphi} ) - \cos 2\varphi \right\} 
\sin^2 \frac{ ( h_{3} - h_{1} ) x }{2}
\biggr]
\nonumber \\ 
&+& 
4 \cos 2\theta_{23} J_{mr} \cos \delta 
\frac{ \Delta_{a} }{ h_{3} - h_{2} } 
\nonumber \\ 
&\times& 
\biggl[
\left( s^2_{13} c^2_{\varphi} - s^2_{\varphi} \right) 
\left\{ \sin^2 \frac{ ( h_{2} - h_{1} ) x }{2} - \sin^2 \frac{ ( h_{3} - h_{1} ) x }{2} \right\}
+ \left\{ c^2_{13} ( 1 + s^2_{\varphi} ) + \cos 2\varphi \right\} 
\sin^2 \frac{ ( h_{3} - h_{2} ) x }{2}
\biggr]
\nonumber \\ 
&-& 
8 c^2_{23} s^2_{23} c^2_{13} s^2_{13} c^2_{\varphi} s^2_{\varphi} 
\cos 2\delta 
%
%\nonumber \\ &\times& 
\frac{ \Delta_{a} }{ h_{3} - h_{1} } 
\biggl[
\sin^2 \frac{ ( h_{2} - h_{1} ) x }{2} 
- \sin^2 \frac{ ( h_{3} - h_{2} ) x }{2} 
+ \sin^2 \frac{ ( h_{3} - h_{1} ) x }{2}
\biggr]
\nonumber \\ 
&-& 
8 c^2_{23} s^2_{23} c^2_{13} s^2_{13} c^2_{\varphi} s^2_{\varphi} 
\cos 2\delta 
%
%\nonumber \\ &\times& 
\frac{ \Delta_{a} }{ h_{3} - h_{2} } 
\biggl[
\sin^2 \frac{ ( h_{2} - h_{1} ) x }{2} 
+ \sin^2 \frac{ ( h_{3} - h_{2} ) x }{2} 
- \sin^2 \frac{ ( h_{3} - h_{1} ) x }{2}
\biggr]
\nonumber \\ 
&+& 
8 J_{mr} \sin \delta 
%%\left\{ 2 s^2_{13} c^2_{\varphi}  + \left( s^2_{13} s^2_{\varphi} - c^2_{\varphi} \right) \right\} 
\left( s^2_{13} - c^2_{13} c^2_{\varphi} \right) 
\frac{ \Delta_{a} }{ h_{3} - h_{1} } 
\sin \frac{ ( h_{3} - h_{2} ) x }{2} 
\sin \frac{ ( h_{2} - h_{1} ) x }{2} 
\sin \frac{ ( h_{1} - h_{3} ) x }{2} 
\nonumber \\ 
&+& 
8 J_{mr} \sin \delta 
%%\left\{ 2 s^2_{13} s^2_{\varphi} + \left( s^2_{13} c^2_{\varphi} - s^2_{\varphi} \right) \right\} 
\left( s^2_{13} - c^2_{13} s^2_{\varphi} \right) 
\frac{ \Delta_{a} }{ h_{3} - h_{2} } 
\sin \frac{ ( h_{3} - h_{2} ) x }{2} 
\sin \frac{ ( h_{2} - h_{1} ) x }{2} 
\sin \frac{ ( h_{1} - h_{3} ) x }{2}. 
\label{P-mutau-1st}
\end{eqnarray}
The unitarity in $\nu_{\mu}$ row at first order 
\begin{eqnarray} 
P(\nu_{\mu} \rightarrow \nu_{e})^{(1)} 
+ P(\nu_{\mu} \rightarrow \nu_{\mu})^{(1)}
+ P(\nu_{\mu} \rightarrow \nu_{\tau})^{(1)} =0
\end{eqnarray}
has been checked.

\section{Parameter degeneracy with the solar-resonance perturbed formula}
\label{sec:parameter-degeneracy}

We use our zeroth-order formulas of neutrino and anti-neutrino oscillation probabilities, $P_{\mu e} \equiv P(\nu_{\mu} \rightarrow \nu_{e})^{(0)}$ and $\bar{P}_{\mu e} \equiv P(\bar{\nu}_{\mu} \rightarrow \bar{\nu}_{e})^{(0)}$, to discuss the parameter degeneracy. We use a simplified setting for treatment of degeneracy, namely, we restrict ourselves to the $\theta_{23} - \delta$ degeneracy which would be left after precision measurement of $\theta_{12}$ and $\theta_{13}$.

We use the notation $y \equiv s^2_{23}$. Then, the set of equations which describe the two degenerate solutions $( y_{1}, \delta_{1} )$ and $( y_{2}, \delta_{2} )$ are given by 
\begin{eqnarray} 
P_{\mu e} 
&=& 
A ( 1 - y_1 ) + B y_1 + C \cos \delta_1 \sin 2\theta_{23} + D \sin \delta_1 \sin 2\theta_{23}, 
\nonumber \\
P_{\mu e} 
&=& 
A ( 1 - y_2 ) + B y_2 + C \cos \delta_2 \sin 2\theta_{23} + D \sin \delta_2 \sin 2\theta_{23}, 
\nonumber \\
\bar{P}_{\mu e} 
&=& 
\bar{A} ( 1 - y_1 ) + \bar{B} y_1 + \bar{C} \cos \delta_1 \sin 2\theta_{23} - \bar{D} \sin \delta_1 \sin 2\theta_{23}, 
\nonumber \\
\bar{P}_{\mu e} 
&=& 
\bar{A} ( 1 - y_2 ) + \bar{B} y_2 + \bar{C} \cos \delta_2 \sin 2\theta_{23} - \bar{D} \sin \delta_2 \sin 2\theta_{23}, 
\label{degeneracy-eq}
\end{eqnarray}
where we have used in (\ref{degeneracy-eq}) the following abbreviated notations 
\begin{eqnarray} 
A &\equiv& 
c^2_{13} \sin^2 2 \varphi 
\sin^2 \frac{ ( h_{2} - h_{1} ) x }{2},
\nonumber \\
B &\equiv& 
\sin^2 2\theta_{13} 
\left[
c^2_{\varphi} \sin^2 \frac{ ( h_{3} - h_{1} ) x }{2} 
+ s^2_{\varphi} \sin^2 \frac{ ( h_{3} - h_{2} ) x }{2} 
- c^2_{\varphi} s^2_{\varphi} \sin^2 \frac{ ( h_{2} - h_{1} ) x }{2} 
\right], 
\nonumber \\
C &\equiv&
2 c^2_{13} s_{13} c_{\varphi} s_{\varphi} 
\left\{
\cos 2 \varphi \sin^2 \frac{ ( h_{2} - h_{1} ) x }{2} 
- \sin^2 \frac{ ( h_{3} - h_{2} ) x }{2} 
+ \sin^2 \frac{ ( h_{3} - h_{1} ) x }{2} 
\right\}, 
\nonumber \\
D &\equiv&
4 c^2_{13} s_{13} c_{\varphi} s_{\varphi} 
\sin \frac{ ( h_{3} - h_{2} ) x }{2} 
\sin \frac{ ( h_{2} - h_{1} ) x }{2} 
\sin \frac{ ( h_{1} - h_{3} ) x }{2}. 
\label{coefficient-def}
\end{eqnarray}
In the degeneracy equation (\ref{degeneracy-eq}), we have made an important approximation that $\left( \sin 2\theta_{23} \right)_{1} = \left( \sin 2\theta_{23} \right)_{2}$. In fact, it is not an approximation, but the first step of iterative procedure used in ref.~\cite{Minakata:2013hgk}. Therefore, $\sin 2\theta_{23}$ in (\ref{degeneracy-eq}) implies $\left( \sin 2\theta_{23} \right)_{1}$. 

We regard $( y_{1}, \delta_{1} )$ as the true solution, and solve (\ref{degeneracy-eq}) to determine a fake solution $( y_{2}, \delta_{2} )$. One can easily show that 
\begin{eqnarray} 
&& 
\sin \delta_2 - \sin \delta_1 = 
\mathcal{S} ( y_2 - y_1 ), 
\nonumber \\
&&
\cos \delta_2 - \cos \delta_1 
= \mathcal{C} ( y_2 - y_1 ),
\label{delta2-solution}
\end{eqnarray}
where 
\begin{eqnarray} 
&& 
\mathcal{S} = 
\frac{1}{ \sin 2\theta_{23} }
\frac{ ( A - B ) \bar{C} - ( \bar{A} - \bar{B} ) C }{ C \bar{D} + \bar{C} D },
\nonumber \\
&&
\mathcal{C} = 
\frac{1}{ \sin 2\theta_{23} }
\frac{ ( A - B ) \bar{D} + ( \bar{A} - \bar{B} ) D }{ C \bar{D} + \bar{C} D }.
\end{eqnarray}
Then, $y_{2}$ is obtained by using $\sin^2 \delta_{2} + \cos^2 \delta_{2} =1$ as 
\begin{eqnarray} 
y_2 = y_1 
- 
\frac{ 2 \left\{ \mathcal{S} \sin \delta_1 + \mathcal{C} \cos \delta_1 \right\} }
{ \left( \mathcal{S}^2 + \mathcal{C}^2 \right) },
\label{y2-solution}
\end{eqnarray}
and $\sin \delta_{2}$ and $\cos \delta_{2}$ are determined by using (\ref{delta2-solution}). In the second iteration, one can use solution of the first step equation, $\left( \sin 2\theta_{23} \right)_{2} \vert_\text{second iteration} = 2 \sqrt{ y_{2} (1-y_{2}) }$. 

Notice that precision measurement of disappearance probabilities $P(\nu_{\mu} \rightarrow \nu_{\mu})$ as well as $P(\nu_{\mu} \rightarrow \nu_{e})$ at around the atmospheric oscillation maximum as planned in T2HK and DUNE would help resolve the $\theta_{23} - \delta$ degeneracy. But, in our ``mathematical'' setting, precision measurement of $P(\nu_{\mu} \rightarrow \nu_{e})$ and its anti-neutrino counterpart at around the solar resonance is sufficient to solve the degeneracy. 

If precision of the measurement of $P(\nu_{\mu} \rightarrow \nu_{e})$ at the solar resonance region and in the usual LBL setting becomes good enough, whose latter is described in ref.~\cite{Minakata:2013hgk}, then the most accurate way of resolving the $\theta_{23} - \delta$ degeneracy would be to combine these two sets of measurements. That is because of $c^2_{23}$ and $s^2_{23}$ dependence of the main oscillation terms of the perturbative formulas around the solar- and atmospheric-resonance regions, respectively. Of course, the quantitative analysis is called for to justify this statement. 

\section{Symmetry of the oscillation probability in ``Simple and Compact'' formulas}
\label{sec:symmetry-MP-formula}

We refer appendix~B in ref.~\cite{Minakata:2015gra} for explicit expressions of the oscillation probabilities in the ``atmospheric resonance'' perturbation theory. It is then easy to recognize that there is a symmetry under the transformations involving angle $\phi$ as (see \cite{Minakata:2015gra} for notations)
\begin{eqnarray} 
&& 
\lambda_{+} \rightarrow \lambda_{-}, 
\hspace{10mm}
\lambda_{-} \rightarrow \lambda_{+}, 
\nonumber \\
&&
c_{\phi} \rightarrow - s_{\phi}, 
\hspace{10mm}
s_{\phi} \rightarrow c_{\phi}, 
\nonumber \\
&&
\cos (\phi - \theta_{12}) \rightarrow - \sin ( \phi - \theta_{12} ), 
\hspace{10mm}
\sin (\phi - \theta_{12}) \rightarrow \cos ( \phi - \theta_{12} ).
\label{phi-transformation}
\end{eqnarray}
which may be summarized as 
\begin{eqnarray} 
&& 
\phi \rightarrow \phi + \frac{\pi}{2}. 
\label{phi-transformation-summary}
\end{eqnarray}
The angle $\phi$ is the matter-dressed $\theta_{13}$ angle, which describes the 1-3 level crossing in matter. The nature of the symmetry is the same as $\varphi \rightarrow \varphi + \frac{\pi}{2}$ symmetry in the ``solar resonance'' perturbation theory, as described in section~\ref{sec:symmetry}.


\begin{thebibliography}{99}

%\cite{Fukuda:1994mc}
\bibitem{Fukuda:1994mc}
  Y.~Fukuda {\it et al.} [Kamiokande Collaboration],
``Atmospheric muon-neutrino / electron-neutrino ratio in the multiGeV energy range,''
  Phys.\ Lett.\ B {\bf 335} (1994) 237.
  doi:10.1016/0370-2693(94)91420-6
  %%CITATION = doi:10.1016/0370-2693(94)91420-6;%%

%\cite{Fukuda:1998mi}
\bibitem{Fukuda:1998mi}
  Y.~Fukuda {\it et al.} [Super-Kamiokande Collaboration],
 ``Evidence for oscillation of atmospheric neutrinos,''
  Phys.\ Rev.\ Lett.\  {\bf 81} (1998) 1562
  doi:10.1103/PhysRevLett.81.1562
  [hep-ex/9807003].
  %%CITATION = doi:10.1103/PhysRevLett.81.1562;%%

%\cite{Eguchi:2002dm}
\bibitem{Eguchi:2002dm}
  K.~Eguchi {\it et al.} [KamLAND Collaboration],
 ``First results from KamLAND: Evidence for reactor anti-neutrino disappearance,''
  Phys.\ Rev.\ Lett.\  {\bf 90} (2003) 021802
  doi:10.1103/PhysRevLett.90.021802
  [hep-ex/0212021].
  %%CITATION = doi:10.1103/PhysRevLett.90.021802;%%

%\cite{Cleveland:1998nv}
\bibitem{Cleveland:1998nv}
  B.~T.~Cleveland, T.~Daily, R.~Davis, Jr., J.~R.~Distel, K.~Lande, C.~K.~Lee, P.~S.~Wildenhain and J.~Ullman,
 ``Measurement of the solar electron neutrino flux with the Homestake chlorine detector,''
  Astrophys.\ J.\  {\bf 496} (1998) 505.
  doi:10.1086/305343
  %%CITATION = doi:10.1086/305343;%%

%\cite{Ahmad:2002jz}
\bibitem{Ahmad:2002jz}
  Q.~R.~Ahmad {\it et al.} [SNO Collaboration],
 ``Direct evidence for neutrino flavor transformation from neutral current interactions in the Sudbury Neutrino Observatory,''
  Phys.\ Rev.\ Lett.\  {\bf 89} (2002) 011301
  doi:10.1103/PhysRevLett.89.011301
  [nucl-ex/0204008].
  %%CITATION = doi:10.1103/PhysRevLett.89.011301;%%

%\cite{Maki:1962mu}
\bibitem{Maki:1962mu}
  Z.~Maki, M.~Nakagawa and S.~Sakata,
``Remarks on the unified model of elementary particles,''
  Prog.\ Theor.\ Phys.\  {\bf 28} (1962) 870.
  doi:10.1143/PTP.28.870
  %%CITATION = doi:10.1143/PTP.28.870;%%

%\cite{Mikheev:1986gs}
\bibitem{Mikheev:1986gs}
  S.~P.~Mikheev and A.~Y.~Smirnov,
``Resonance Amplification of Oscillations in Matter and Spectroscopy of Solar Neutrinos,''
  Sov.\ J.\ Nucl.\ Phys.\  {\bf 42} (1985) 913
   [Yad.\ Fiz.\  {\bf 42} (1985) 1441].
  %%CITATION = SJNCA,42,913;%%

%\cite{Wolfenstein:1977ue}
\bibitem{Wolfenstein:1977ue}
  L.~Wolfenstein,
``Neutrino Oscillations in Matter,''
  Phys.\ Rev.\ D {\bf 17} (1978) 2369.
  doi:10.1103/PhysRevD.17.2369
  %%CITATION = doi:10.1103/PhysRevD.17.2369;%%

%\cite{Gando:2013nba} 
\bibitem{Gando:2013nba}
  A.~Gando {\it et al.} [KamLAND Collaboration],
``Reactor On-Off Antineutrino Measurement with KamLAND,''
  Phys.\ Rev.\ D {\bf 88} (2013) no.3,  033001
  doi:10.1103/PhysRevD.88.033001
  [arXiv:1303.4667 [hep-ex]].
  %%CITATION = doi:10.1103/PhysRevD.88.033001;%%

%\cite{Abe:2018wpn} 
\bibitem{Abe:2018wpn}
  K.~Abe {\it et al.} [T2K Collaboration],
 ``Search for CP Violation in Neutrino and Antineutrino Oscillations by the T2K Experiment with $2.2\times10^{21}$ Protons on Target,''
  Phys.\ Rev.\ Lett.\  {\bf 121} (2018) no.17,  171802
  doi:10.1103/PhysRevLett.121.171802
  [arXiv:1807.07891 [hep-ex]].
  %%CITATION = doi:10.1103/PhysRevLett.121.171802;%%

%\cite{Abe:2017bay} 
\bibitem{Abe:2017bay}
  K.~Abe {\it et al.} [T2K Collaboration],
``Updated T2K measurements of muon neutrino and antineutrino disappearance using 1.5$\times$10$^{21}$ protons on target,''
  Phys.\ Rev.\ D {\bf 96} (2017) no.1,  011102
  doi:10.1103/PhysRevD.96.011102
  [arXiv:1704.06409 [hep-ex]].
  %%CITATION = doi:10.1103/PhysRevD.96.011102;%%

%\cite{NOvA:2018gge} 
\bibitem{NOvA:2018gge}
  M.~A.~Acero {\it et al.} [NOvA Collaboration],
``New constraints on oscillation parameters from $\nu_e$ appearance and $\nu_\mu$ disappearance in the NOvA experiment,''
  Phys.\ Rev.\ D {\bf 98} (2018) 032012
  doi:10.1103/PhysRevD.98.032012
  [arXiv:1806.00096 [hep-ex]].
  %%CITATION = doi:10.1103/PhysRevD.98.032012;%%

%\cite{Jiang:2019xwn} 
\bibitem{Jiang:2019xwn}
  M.~Jiang {\it et al.} [Super-Kamiokande Collaboration],
``Atmospheric Neutrino Oscillation Analysis With Improved Event Reconstruction in Super-Kamiokande IV,''
  arXiv:1901.03230 [hep-ex].
  %%CITATION = ARXIV:1901.03230;%%


%\cite{Adey:2018zwh} 
\bibitem{Adey:2018zwh}
  D.~Adey {\it et al.} [Daya Bay Collaboration],
``Measurement of the Electron Antineutrino Oscillation with 1958 Days of Operation at Daya Bay,''
  Phys.\ Rev.\ Lett.\  {\bf 121} (2018) no.24,  241805
  doi:10.1103/PhysRevLett.121.241805
  [arXiv:1809.02261 [hep-ex]].
  %%CITATION = doi:10.1103/PhysRevLett.121.241805;%%

%\cite{Bak:2018ydk}
\bibitem{Bak:2018ydk}
  G.~Bak {\it et al.} [RENO Collaboration],
  %``Measurement of Reactor Antineutrino Oscillation Amplitude and Frequency at RENO,''
  Phys.\ Rev.\ Lett.\  {\bf 121} (2018) no.20,  201801
  doi:10.1103/PhysRevLett.121.201801
  [arXiv:1806.00248 [hep-ex]].
  %%CITATION = doi:10.1103/PhysRevLett.121.201801;%%

%\cite{Nakano}
\bibitem{Nakano}
  Y.~Nakano,
``$^8$B solar neutrino spectrum measurement using Super-Kamiokande IV'', Ph.D. thesis, University of Tokyo, December 2015.
  %%CITATION = INSPIRE-1667166;%%

%\cite{Okumura-ICRR-seminar}
\bibitem{Okumura-ICRR-seminar}
K.~Okumura (for the T2K Collaboration), 
``Updated results from T2K experiment with $3.13 \times 10^{21}$ POT, ICRR Seminar, January 10th, 2019. 

%\cite{Abe:2015zbg} 
\bibitem{Abe:2015zbg}
  K.~Abe {\it et al.} [Hyper-Kamiokande Proto- Collaboration],
``Physics potential of a long-baseline neutrino oscillation experiment using a J-PARC neutrino beam and Hyper-Kamiokande,''
  PTEP {\bf 2015} (2015) 053C02
  doi:10.1093/ptep/ptv061
  [arXiv:1502.05199 [hep-ex]].
  %%CITATION = doi:10.1093/ptep/ptv061;%%

%\cite{Acciarri:2015uup} 
\bibitem{Acciarri:2015uup}
  R.~Acciarri {\it et al.} [DUNE Collaboration],
 ``Long-Baseline Neutrino Facility (LBNF) and Deep Underground Neutrino Experiment (DUNE) : Volume 2: The Physics Program for DUNE at LBNF,''
  arXiv:1512.06148 [physics.ins-det].
  %%CITATION = ARXIV:1512.06148;%%

%\cite{Minakata:2001qm}
\bibitem{Minakata:2001qm}
  H.~Minakata and H.~Nunokawa,
``Exploring neutrino mixing with low-energy superbeams,''
  JHEP {\bf 0110} (2001) 001
  doi:10.1088/1126-6708/2001/10/001
  [hep-ph/0108085].
  %%CITATION = doi:10.1088/1126-6708/2001/10/001;%%

%\cite{Abe:2016ero}
\bibitem{Abe:2016ero}
 K.~Abe {\it et al.} [Hyper-Kamiokande Collaboration],
``Physics potentials with the second Hyper-Kamiokande detector in Korea,''
  PTEP {\bf 2018} (2018) no.6,  063C01
  doi:10.1093/ptep/pty044
  [arXiv:1611.06118 [hep-ex]].
  %%CITATION = doi:10.1093/ptep/pty044;%%

%\cite{TheIceCube-Gen2:2016cap}
\bibitem{TheIceCube-Gen2:2016cap}
  M.~G.~Aartsen {\it et al.} [IceCube Collaboration],
 ``PINGU: A Vision for Neutrino and Particle Physics at the South Pole,''
  J.\ Phys.\ G {\bf 44} (2017) no.5,  054006
  doi:10.1088/1361-6471/44/5/054006
  [arXiv:1607.02671 [hep-ex]].
  %%CITATION = doi:10.1088/1361-6471/44/5/054006;%%

%\cite{Adrian-Martinez:2016zzs}
\bibitem{Adrian-Martinez:2016zzs}
  S.~Adrián-Martínez {\it et al.},
``Intrinsic limits on resolutions in muon- and electron-neutrino charged-current events in the KM3NeT/ORCA detector,''
  JHEP {\bf 1705} (2017) 008
  doi:10.1007/JHEP05(2017)008
  [arXiv:1612.05621 [physics.ins-det]].
  %%CITATION = doi:10.1007/JHEP05(2017)008;%%

%\cite{An:2015jdp}
\bibitem{An:2015jdp}
  F.~An {\it et al.} [JUNO Collaboration],
``Neutrino Physics with JUNO,''
  J.\ Phys.\ G {\bf 43} (2016) no.3,  030401
  doi:10.1088/0954-3899/43/3/030401
  [arXiv:1507.05613 [physics.ins-det]].
  %%CITATION = doi:10.1088/0954-3899/43/3/030401;%%

%\cite{Minakata-talk}
\bibitem{Minakata-talk}
H.~Minakata, ``Terrestrial solar-scale neutrino oscillations'', 
Talk at Workshop for Atmospheric Neutrino Production, March 20-22, 2019, Nagoya University, Nagoya, Japan.

%\cite{Peres:2003wd} 
\bibitem{Peres:2003wd}
  O.~L.~G.~Peres and A.~Y.~Smirnov,
``Atmospheric neutrinos: LMA oscillations, U(e3) induced interference and CP violation,''
  Nucl.\ Phys.\ B {\bf 680} (2004) 479
  doi:10.1016/j.nuclphysb.2003.12.017
  [hep-ph/0309312].
  %%CITATION = doi:10.1016/j.nuclphysb.2003.12.017;%%

%\cite{Peres:2009xe}
\bibitem{Peres:2009xe}
  O.~L.~G.~Peres and A.~Y.~Smirnov,
``Oscillations of very low energy atmospheric neutrinos,''
  Phys.\ Rev.\ D {\bf 79} (2009) 113002
  doi:10.1103/PhysRevD.79.113002
  [arXiv:0903.5323 [hep-ph]].
  %%CITATION = doi:10.1103/PhysRevD.79.113002;%%

%\cite{Akhmedov:2008qt} 
\bibitem{Akhmedov:2008qt} 
  E.~K.~Akhmedov, M.~Maltoni and A.~Y.~Smirnov,
 ``Neutrino oscillograms of the Earth: Effects of 1-2 mixing and CP-violation,''
  JHEP {\bf 0806}, 072 (2008)
  doi:10.1088/1126-6708/2008/06/072
  [arXiv:0804.1466 [hep-ph]].
  %%CITATION = doi:10.1088/1126-6708/2008/06/072;%%

%\cite{Razzaque:2014vba}
\bibitem{Razzaque:2014vba}
  S.~Razzaque and A.~Y.~Smirnov,
 ``Super-PINGU for measurement of the leptonic CP-phase with atmospheric neutrinos,''
  JHEP {\bf 1505} (2015) 139
  doi:10.1007/JHEP05(2015)139
  [arXiv:1406.1407 [hep-ph]].
  %%CITATION = doi:10.1007/JHEP05(2015)139;%%

%\cite{Mari-etal-Nutele2019}
\bibitem{Mari-etal-Nutele2019}
S.~M.~Mari, C.~Martellini, P.~Montini, G.~Settanta, 
``Atmospheric neutrino spectrum reconstruction with JUNO'', 
Talk at XVIII International Workshop on Neutrino Telescopes, March 18-22, 2019, Venice, Italy.

%\cite{Kelly:2019itm}
\bibitem{Kelly:2019itm}
  K.~J.~Kelly, P.~A.~Machado, I.~Martinez Soler, S.~J.~Parke and Y.~F.~Perez Gonzalez,
``Sub-GeV Atmospheric Neutrinos and CP-Violation in DUNE,''
  arXiv:1904.02751 [hep-ph].
  %%CITATION = ARXIV:1904.02751;%%

%\cite{Minakata:2000ee}
\bibitem{Minakata:2000ee}
  H.~Minakata and H.~Nunokawa,
 ``Measuring leptonic CP violation by low-energy neutrino oscillation experiments,''
  Phys.\ Lett.\ B {\bf 495} (2000) 369
  doi:10.1016/S0370-2693(00)01249-1
  [hep-ph/0004114].
  %%CITATION = doi:10.1016/S0370-2693(00)01249-1;%%


%\cite{Asano:2011nj}
\bibitem{Asano:2011nj}
  K.~Asano and H.~Minakata,
``Large-Theta(13) Perturbation Theory of Neutrino Oscillation for Long-Baseline Experiments,''
  JHEP {\bf 1106} (2011) 022
  doi:10.1007/JHEP06(2011)022
  [arXiv:1103.4387 [hep-ph]].
  %%CITATION = doi:10.1007/JHEP06(2011)022;%%


%\cite{Arafune:1997hd}
\bibitem{Arafune:1997hd}
  J.~Arafune, M.~Koike and J.~Sato,
``CP violation and matter effect in long baseline neutrino oscillation experiments,''
  Phys.\ Rev.\ D {\bf 56} (1997) 3093
   Erratum: [Phys.\ Rev.\ D {\bf 60} (1999) 119905]
  doi:10.1103/PhysRevD.60.119905, 10.1103/PhysRevD.56.3093
  [hep-ph/9703351].
  %%CITATION = doi:10.1103/PhysRevD.60.119905, 10.1103/PhysRevD.56.3093;%%

%\cite{Cervera:2000kp}
\bibitem{Cervera:2000kp}
  A.~Cervera, A.~Donini, M.~B.~Gavela, J.~J.~Gomez Cadenas, P.~Hernandez, O.~Mena and S.~Rigolin,
 ``Golden measurements at a neutrino factory,''
  Nucl.\ Phys.\ B {\bf 579} (2000) 17
   Erratum: [Nucl.\ Phys.\ B {\bf 593} (2001) 731]
  doi:10.1016/S0550-3213(00)00606-4, 10.1016/S0550-3213(00)00221-2
  [hep-ph/0002108].
  %%CITATION = doi:10.1016/S0550-3213(00)00606-4, 10.1016/S0550-3213(00)00221-2;%%

%\cite{Kikuchi:2008vq}
\bibitem{Kikuchi:2008vq}
  T.~Kikuchi, H.~Minakata and S.~Uchinami,
``Perturbation Theory of Neutrino Oscillation with Nonstandard Neutrino Interactions,''
  JHEP {\bf 0903} (2009) 114
  doi:10.1088/1126-6708/2009/03/114
  [arXiv:0809.3312 [hep-ph]].
  %%CITATION = doi:10.1088/1126-6708/2009/03/114;%%

%\cite{Martinez-Soler:2018lcy}
\bibitem{Martinez-Soler:2018lcy}
  I.~Martinez-Soler and H.~Minakata,
``Standard versus Non-Standard CP Phases in Neutrino Oscillation in Matter with Non-Unitarity,''
  arXiv:1806.10152 [hep-ph].
  %%CITATION = ARXIV:1806.10152;%%

%\cite{BurguetCastell:2001ez}
\bibitem{BurguetCastell:2001ez}
  J.~Burguet-Castell, M.~B.~Gavela, J.~J.~Gomez-Cadenas, P.~Hernandez and O.~Mena,
``On the Measurement of leptonic CP violation,''
  Nucl.\ Phys.\ B {\bf 608} (2001) 301
  doi:10.1016/S0550-3213(01)00248-6
  [hep-ph/0103258].
  %%CITATION = doi:10.1016/S0550-3213(01)00248-6;%%


%\cite{Fogli:1996pv}
\bibitem{Fogli:1996pv}
  G.~L.~Fogli and E.~Lisi,
``Tests of three flavor mixing in long baseline neutrino oscillation experiments,''
  Phys.\ Rev.\ D {\bf 54} (1996) 3667
  doi:10.1103/PhysRevD.54.3667
  [hep-ph/9604415].
  %%CITATION = doi:10.1103/PhysRevD.54.3667;%%

%\cite{Kimura:2002wd}
\bibitem{Kimura:2002wd}
  K.~Kimura, A.~Takamura and H.~Yokomakura,
``Exact formulas and simple CP dependence of neutrino oscillation probabilities in matter with constant density,''
  Phys.\ Rev.\ D {\bf 66} (2002) 073005
  doi:10.1103/PhysRevD.66.073005
  [hep-ph/0205295].
  %%CITATION = doi:10.1103/PhysRevD.66.073005;%%

%\cite{Zaglauer:1988gz}
\bibitem{Zaglauer:1988gz}
  H.~W.~Zaglauer and K.~H.~Schwarzer,
``The Mixing Angles in Matter for Three Generations of Neutrinos and the Msw Mechanism,''
  Z.\ Phys.\ C {\bf 40} (1988) 273.
  doi:10.1007/BF01555889
  %%CITATION = doi:10.1007/BF01555889;%%

%\cite{Tanabashi:2018oca} 
\bibitem{Tanabashi:2018oca}
  M.~Tanabashi {\it et al.} [Particle Data Group],
``Review of Particle Physics,''
  Phys.\ Rev.\ D {\bf 98} (2018) no.3,  030001.
  doi:10.1103/PhysRevD.98.030001
  %%CITATION = doi:10.1103/PhysRevD.98.030001;%%

%\cite{Esteban:2018azc} 
\bibitem{Esteban:2018azc}
  I.~Esteban, M.~C.~Gonzalez-Garcia, A.~Hernandez-Cabezudo, M.~Maltoni and T.~Schwetz,
``Global analysis of three-flavour neutrino oscillations: synergies and tensions in the determination of $\theta_{23}, \delta_{CP}$, and the mass ordering,''
  JHEP {\bf 1901} (2019) 106
  doi:10.1007/JHEP01(2019)106
  [arXiv:1811.05487 [hep-ph]].
  %%CITATION = doi:10.1007/JHEP01(2019)106;%%

%\cite{Capozzi:2018ubv} 
\bibitem{Capozzi:2018ubv}
  F.~Capozzi, E.~Lisi, A.~Marrone and A.~Palazzo,
``Current unknowns in the three neutrino framework,''
  Prog.\ Part.\ Nucl.\ Phys.\  {\bf 102} (2018) 48
  doi:10.1016/j.ppnp.2018.05.005
  [arXiv:1804.09678 [hep-ph]].
  %%CITATION = doi:10.1016/j.ppnp.2018.05.005;%%

%\cite{deSalas:2017kay}
\bibitem{deSalas:2017kay}
  P.~F.~de Salas, D.~V.~Forero, C.~A.~Ternes, M.~Tortola and J.~W.~F.~Valle,
``Status of neutrino oscillations 2018: 3$\sigma$ hint for normal mass ordering and improved CP sensitivity,''
  Phys.\ Lett.\ B {\bf 782} (2018) 633
  doi:10.1016/j.physletb.2018.06.019
  [arXiv:1708.01186 [hep-ph]].
  %%CITATION = doi:10.1016/j.physletb.2018.06.019;%%


%\cite{Ohlsson:2012kf} 
\bibitem{Ohlsson:2012kf}
  T.~Ohlsson,
 ``Status of non-standard neutrino interactions,''
  Rept.\ Prog.\ Phys.\  {\bf 76} (2013) 044201
  doi:10.1088/0034-4885/76/4/044201
  [arXiv:1209.2710 [hep-ph]].
  %%CITATION = doi:10.1088/0034-4885/76/4/044201;%%

%\cite{Miranda:2015dra}
\bibitem{Miranda:2015dra}
  O.~G.~Miranda and H.~Nunokawa,
``Non standard neutrino interactions: current status and future prospects,''
  New J.\ Phys.\  {\bf 17} (2015) no.9,  095002
  doi:10.1088/1367-2630/17/9/095002
  [arXiv:1505.06254 [hep-ph]].
  %%CITATION = doi:10.1088/1367-2630/17/9/095002;%%

%\cite{Antusch:2006vwa} 
\bibitem{Antusch:2006vwa}
  S.~Antusch, C.~Biggio, E.~Fernandez-Martinez, M.~B.~Gavela and J.~Lopez-Pavon,
 ``Unitarity of the Leptonic Mixing Matrix,''
  JHEP {\bf 0610} (2006) 084
  doi:10.1088/1126-6708/2006/10/084
  [hep-ph/0607020].
  %%CITATION = doi:10.1088/1126-6708/2006/10/084;%%

%\cite{Escrihuela:2015wra}
\bibitem{Escrihuela:2015wra}
  F.~J.~Escrihuela, D.~V.~Forero, O.~G.~Miranda, M.~Tórtola and J.~W.~F.~Valle,
``On the description of non-unitary neutrino mixing,''
  Phys.\ Rev.\ D {\bf 92} (2015) no.5,  053009
  doi:10.1103/PhysRevD.92.053009
  [arXiv:1503.08879 [hep-ph]].
  %%CITATION = doi:10.1103/PhysRevD.92.053009;%%

%\cite{Fong:2016yyh} 
\bibitem{Fong:2016yyh}
  C.~S.~Fong, H.~Minakata and H.~Nunokawa,
 ``A framework for testing leptonic unitarity by neutrino oscillation experiments,''
  JHEP {\bf 1702} (2017) 114
  doi:10.1007/JHEP02(2017)114
  [arXiv:1609.08623 [hep-ph]].
  %%CITATION = doi:10.1007/JHEP02(2017)114;%%

%\cite{Fong:2017gke}
\bibitem{Fong:2017gke}
  C.~S.~Fong, H.~Minakata and H.~Nunokawa,
``Non-unitary evolution of neutrinos in matter and the leptonic unitarity test,''
  JHEP {\bf 1902} (2019) 015
  doi:10.1007/JHEP02(2019)015
  [arXiv:1712.02798 [hep-ph]].
  %%CITATION = doi:10.1007/JHEP02(2019)015;%%

%\cite{Minakata:2015gra}
\bibitem{Minakata:2015gra}
  H.~Minakata and S.~J.~Parke,
``Simple and Compact Expressions for Neutrino Oscillation Probabilities in Matter,''
  JHEP {\bf 1601} (2016) 180
  doi:10.1007/JHEP01(2016)180
  [arXiv:1505.01826 [hep-ph]].
  %%CITATION = doi:10.1007/JHEP01(2016)180;%%

%\cite{Agarwalla:2013tza}
\bibitem{Agarwalla:2013tza}
  S.~K.~Agarwalla, Y.~Kao and T.~Takeuchi,
``Analytical approximation of the neutrino oscillation matter effects at large $\theta_{13}$,''
  JHEP {\bf 1404} (2014) 047
  doi:10.1007/JHEP04(2014)047
  [arXiv:1302.6773 [hep-ph]].
  %%CITATION = doi:10.1007/JHEP04(2014)047;%%

%\cite{Denton:2016wmg}
\bibitem{Denton:2016wmg}
  P.~B.~Denton, H.~Minakata and S.~J.~Parke,
``Compact Perturbative Expressions For Neutrino Oscillations in Matter,''
  JHEP {\bf 1606} (2016) 051
  doi:10.1007/JHEP06(2016)051
  [arXiv:1604.08167 [hep-ph]].
  %%CITATION = doi:10.1007/JHEP06(2016)051;%%

%\cite{Arafune:1996bt} 
\bibitem{Arafune:1996bt}
  J.~Arafune and J.~Sato,
``CP and T violation test in neutrino oscillation,''
  Phys.\ Rev.\ D {\bf 55} (1997) 1653
  doi:10.1103/PhysRevD.55.1653
  [hep-ph/9607437].
  %%CITATION = doi:10.1103/PhysRevD.55.1653;%%

%\cite{Freund:2001pn}
\bibitem{Freund:2001pn}
  M.~Freund,
``Analytic approximations for three neutrino oscillation parameters and probabilities in matter,''
  Phys.\ Rev.\ D {\bf 64} (2001) 053003
  doi:10.1103/PhysRevD.64.053003
  [hep-ph/0103300].
  %%CITATION = doi:10.1103/PhysRevD.64.053003;%%

%\cite{Akhmedov:2004ny}
\bibitem{Akhmedov:2004ny}
  E.~K.~Akhmedov, R.~Johansson, M.~Lindner, T.~Ohlsson and T.~Schwetz,
``Series expansions for three flavor neutrino oscillation probabilities in matter,''
  JHEP {\bf 0404} (2004) 078
  doi:10.1088/1126-6708/2004/04/078
  [hep-ph/0402175].
  %%CITATION = doi:10.1088/1126-6708/2004/04/078;%%

%\cite{Ishitsuka:2005qi} 
\bibitem{Ishitsuka:2005qi}
  M.~Ishitsuka, T.~Kajita, H.~Minakata and H.~Nunokawa,
``Resolving neutrino mass hierarchy and CP degeneracy by two identical detectors with different baselines,''
  Phys.\ Rev.\ D {\bf 72} (2005) 033003
  doi:10.1103/PhysRevD.72.033003
  [hep-ph/0504026].
  %%CITATION = doi:10.1103/PhysRevD.72.033003;%%

%\cite{Baussan:2013zcy}
\bibitem{Baussan:2013zcy}
  E.~Baussan {\it et al.} [ESSnuSB Collaboration],
``A very intense neutrino super beam experiment for leptonic CP violation discovery based on the European spallation source linac,''
  Nucl.\ Phys.\ B {\bf 885} (2014) 127
  doi:10.1016/j.nuclphysb.2014.05.016
  [arXiv:1309.7022 [hep-ex]].
  %%CITATION = doi:10.1016/j.nuclphysb.2014.05.016;%%

%\cite{Nunokawa-private}
\bibitem{Nunokawa-private}
  H.~Nunokawa, private communications.

%\cite{Minakata:2013hgk}
\bibitem{Minakata:2013hgk}
  H.~Minakata and S.~J.~Parke,
``Correlated, precision measurements of $\theta_{23}$ and $\delta$ using only the electron neutrino appearance experiments,''
  Phys.\ Rev.\ D {\bf 87} (2013) no.11,  113005
  doi:10.1103/PhysRevD.87.113005
  [arXiv:1303.6178 [hep-ph]].
  %%CITATION = doi:10.1103/PhysRevD.87.113005;%%

%\cite{Kumar:2017sdq}
\bibitem{Kumar:2017sdq}
  S.~Ahmed {\it et al.} [ICAL Collaboration],
``Physics Potential of the ICAL detector at the India-based Neutrino Observatory (INO),''
  Pramana {\bf 88} (2017) no.5,  79
  doi:10.1007/s12043-017-1373-4
  [arXiv:1505.07380 [physics.ins-det]].
  %%CITATION = doi:10.1007/s12043-017-1373-4;%%

%\cite{Okumura-private}
\bibitem{Okumura-private}
K.~Okumura, private communications.

%\cite{Minakata:2002jv}
\bibitem{Minakata:2002jv}
  H.~Minakata, H.~Sugiyama, O.~Yasuda, K.~Inoue and F.~Suekane,
``Reactor measurement of theta(13) and its complementarity to long baseline experiments,''
  Phys.\ Rev.\ D {\bf 68} (2003) 033017
   Erratum: [Phys.\ Rev.\ D {\bf 70} (2004) 059901]
  doi:10.1103/PhysRevD.70.059901, 10.1103/PhysRevD.68.033017
  [hep-ph/0211111].
  %%CITATION = doi:10.1103/PhysRevD.70.059901, 10.1103/PhysRevD.68.033017;%%




\end{thebibliography}
\end{document}